\documentclass[usenatbib,useAMS]{mn2e}
\usepackage{times}
\usepackage{epsfig}
\usepackage{amsmath}
\usepackage{natbib}
\usepackage{longtable}
\title[NIR images of H$\delta-$strong galaxies]{Near-infrared imaging of 222 nearby H$\delta$-strong galaxies from the SDSS$^1$}
\author[Balogh et al.]{Michael L. Balogh$^{2,3}$, Chris Miller$^{4}$,Robert
  Nichol$^{5}$, Ann Zabludoff$^{6}$ \& Tomo Goto$^{7}$\\
$^{1}$Based on observations made with the United
  Kingdom Infrared Telescope\\
$^{2}$Department of Physics, University of Waterloo,
  Waterloo, ON, Canada N2L 3G1\\
$^{3}$E-mail: mbalogh@uwaterloo.ca\\
$^{4}$Cerro-Tololo Inter-American Observatory, NOAO, Casilla 603, La Serena, Chile\\
$^{5}$Institute of Cosmology and Gravitation, Mercantile
  House, Hampshire Terrace, University of Portsmouth, Portsmouth, UK PO1 2EG\\
$^{6}$Steward Observatory, University of Arizona Tucson, AZ, 85721 USA\\
$^{7}$Space Telescope Science Institute, 3700 San Martin,  Baltimore,
MD 21218\\
}
\date{\today}
\def\gtrsim{\mathrel{\raise0.35ex\hbox{$\scriptstyle >$}\kern-0.6em
\lower0.40ex\hbox{{$\scriptstyle \sim$}}}}
\def\lesssim{\mathrel{\raise0.35ex\hbox{$\scriptstyle <$}\kern-0.6em
\lower0.40ex\hbox{{$\scriptstyle \sim$}}}}

\def\ewhd{{\,\rm W_\circ(H\delta)}}
\def\ewha{{\,\rm W_\circ(H\alpha)}}
\def\oii{[OII]}
\def\ewoii{{\,\rm W_\circ([O{\sc II}])}}

\begin{document}
\maketitle
\begin{abstract}
We present UFTI K-band imaging observations of 222 galaxies that are selected
from the Sloan Digital Sky Survey to have unusually strong H$\delta$
absorption equivalent widths, $\ewhd>4$\AA.  Using {\sc GIM2D}, the images are fit with
two--dimensional surface brightness models consisting of a simple disk
and bulge component to derive
the fraction of luminosity in the bulge, $B/T$.  We find that the
galaxies with weak or absent H$\alpha$ or [O{\sc ii}]$\lambda$3727
emission (known as k+a galaxies) are predominantly bulge--dominated (with a mode
of $B/T\sim 0.6$), while
galaxies with nebular emission (known as e[a] galaxies) are mostly
disk--dominated ($B/T\sim 0.1$).  The morphologies and $(r-k)$ colours
of most k+a galaxies are inconsistent with the hypothesis that they result from the
truncation of star formation in normal, spiral galaxies.
However, their $(u-g)$ and $(r-k)$ colours, as
well as their H$\delta$ line strengths, form a
sequence that is well matched by a model
in which $>5$ per cent of the stellar mass has been produced in a recent
starburst.   The lack of scatter in the dust-sensitive $(r-k)$ colours
suggests that the unusual spectra of k+a galaxies  are not
  due to the effects of dust.  The e(a) galaxies, on the other hand, have a colour
distribution that is distinct from the k+a population, and typical
of normal or dusty ($\tau_V\sim2$) spiral galaxies.  We conclude that
many e(a) galaxies are not progenitors of k+a galaxies, but are a separate phenomenon.
Both k+a and e(a) galaxies reside in environments (characterized by the local
density of galaxies brighter than $M_r=-20$) that are typical of normal
galaxies and that are inconsistent with overdense regions like rich galaxy
clusters.
\end{abstract}
\section{Introduction}
A possible interpretation of the diverse properties of galaxies is
that, under the right circumstances, they can transform from one
type to another \citep[e.g.][]{Baldry03}.  For example, galaxy mergers are an effective way to
transform gas--rich, star--forming spiral galaxies into gas--poor,
passively evolving elliptical galaxies \citep{Barnes,Springel}.
Support for this interpretation is best obtained by
finding examples of galaxies in the process of transformation;
however, these examples will be rare if the transformation timescale is short.

One of the best candidates for such a transition population is the class of galaxies
with spectra that have unusually strong Balmer--line absorption, but
that lack emission
lines.  Although originally discovered in galaxy clusters at $z\sim 0.3$
\citep{DG82}, later work showed that these galaxies are also found in
low--density environments, at low redshift \citep{Z+96,SDSS_Coma_short,tomo-EA1,Q+04}
and at higher redshift \citep{D+99,PSG}.  Stellar population
synthesis modelling suggests that the spectra are best modelled as
a post-starburst, with all star formation ceasing following the burst
\citep{CS87,DG92,Barger,P+99,PSG}. Less extreme examples (i.e. with
weaker Balmer absorption lines) do not necessarily require a burst, but
can be modelled as a normal star--forming galaxy in which star
formation is suddenly truncated.  Since the lifetime of the enhanced
Balmer lines is short ($\lesssim 0.5$ Gyr), even a small observed
population of such galaxies might indicate that a significant fraction
of all galaxies have undergone a transformation via this phase; it is
even possible that they represent the route by which all early--type
galaxies form \citep{N+01,TranEA}.  

The spectral properties (e.g. line strengths) of these unusual galaxies are not completely
disjoint from those of the normal galaxy population, but rather represent the
tail of a continuous distribution \citep[e.g.][]{Z+96,PSG}.  Given the large number of free parameters available to model the
spectra (e.g. metallicity, initial mass function, and multi-component
dust models, in addition to the parametrization of the star formation
history), it is not therefore possible to uniquely define the
spectral characteristics of a post--starburst galaxy,
and this has led to a variety of definitions and nomenclatures.  We
will adopt the nomenclature (but not the precise definition) of
\citet{D+99}, and call galaxies with strong Balmer absorption, but no
detectable nebular emission, k+a galaxies (because the spectrum
approximately resembles a combination of k-star and a-star spectra).
Galaxies with strong Balmer lines and some emission we will
refer to as e(a).  Furthermore, for clarity, when referring to previous work we will
retain this nomenclature, although the definitions vary significantly
between samples.  In particular, the use of the H$\alpha$ emission
line, often not available in earlier spectroscopic samples, makes a
significant difference to the sample selection \citep[e.g.][]{A2390,Q+04,BlakeEA}.
Our precise definitions of the k+a and e(a) galaxies are given in \S~\ref{sec-sample}.

The spectra of  e(a) galaxies are even more difficult to interpret than
those of k+a galaxies, as it is  difficult to
find a model that predicts both emission lines (arising from star
formation rather than an active nucleus) and strong Balmer
absorption.  This is primarily because the OB stars, which dominate
the optical luminosity and are required to
produce the emission lines, have intrinsically weak Balmer absorption
lines.  Thus, the relationship between k+a and e(a) galaxies is
unknown.  One possible model is that e(a) galaxies have a two--phase
dust distribution, in which the OB stars are preferentially obscured,
relative to the A-stars \citep{PW00}.  It has been
suggested \citep{P+99} that some e(a) galaxies may be  in the midst of a
dust--obscured starburst and will resemble k+a galaxies when the star
formation ends.  The fact that some merging starburst systems and
ultraluminous infrared galaxies exhibit
e(a) spectra may support this hypothesis \citep{LK}.  Another possibility is that some k+a galaxies are
extreme examples of the e(a) phenomenon, where the dust obscuration is
strong enough to completely eliminate the emission lines
\citep{Smail-radio,A2390_BM}. However, this latter interpretation is not likely the
case for the majority of the k+a population, which are undetected in HI and
radio continuum \citep{Chang01,MO_ka}.

Until recently, detailed studies of k+a and e(a) galaxies have been
limited to small samples \citep[e.g.][]{Z+96,Galaz}.  
Long-slit spectroscopy \citep{N+01} and high-resolution {\it Hubble
  Space Telescope} imaging \citep{YZZ+} of some k+a galaxies suggest
that they result from the recent merger of gas--rich disk galaxies and
will evolve into relaxed spheroidal galaxies that lie on the
fundamental plane.
However, because of the small sample size it is not clear how
general the interpretation is \citep[e.g.][]{Galaz,Caldwell,Bart01,TranEA}.
Recently, large samples of nearby k+a and e(a) galaxies have been compiled from
the Sloan Digital Sky Survey \citep{tomo-EA1,Q+04} and the 2dF Galaxy
Redshift Survey \citep{BlakeEA}.  These studies
confirm the extreme rarity of such galaxies; for example,
\citet{tomo-EA1} find that $<0.1$ per cent of all galaxies satisfy our
definition of a k+a galaxy.   
\citet{Q+04} and \citet{BlakeEA} find that most of these galaxies are bulge--dominated and
are found in all environments.  However, the morphologies are based on
visual light, which is sensitive to recent star formation.  In the
present paper, we present near-infrared observations for a large sample of
k+a and e(a) galaxies drawn from the \citet{tomo-EA1} sample to
directly study
the stellar mass and its morphological distribution.

The paper is structured as follows.  In \S~\ref{sec-data} we describe
the galaxy sample and the photometric and morphological measurements.
The main results are presented in \S~\ref{sec-results}, and the
colours, H$\delta$ line strengths and luminosities are compared with
various model predictions in \S~\ref{sec-interp}.  Our conclusions
regarding the connection between e(a) and k+a galaxies, as well as a
comparison with other results in the literature, are given in  \S~\ref{sec-discuss}.  Finally we summarize our
findings in \S~\ref{sec-conc}.
For cosmology--dependent quantities, we assume a matter density
$\Omega_m=0.3$, a dark energy component $\Omega_\Lambda=0.7$ and a
Hubble constant of 70~km~s$^{-1}$~Mpc$^{-1}$.

\section{Data}\label{sec-data}
\subsection{The Galaxy Sample}\label{sec-sample}
We selected our sample from the catalogue of \citet{tomo-EA1}, based
on the first data release of the Sloan Digital Sky Survey \citep{DR1}.
This catalogue contains all galaxies for which the rest frame
equivalent width of the H$\delta$ absorption line is $\ewhd>4$\AA\ (with
2$\sigma$ confidence), as
measured from the spectrum using a Gaussian--fitting technique.  A
second Gaussian was fit to correct for
emission--filling when emission lines are present.  From this sample, we identify two main types of
galaxies: e(a) galaxies have either H$\alpha$ or \oii\ detected in
emission with at least 2$\sigma$ confidence (but see below), while for k+a galaxies
these emission lines are undetected at the same significance level.
These definitions are similar in spirit to those of \citet{D+99}.
However, our definition of a k+a galaxy is stricter in the sense that we
require both stronger absorption in H$\delta$ (4\AA\ compared with
3\AA) and our data quality and wavelength coverage allows us
to be more comprehensive in excluding galaxies with emission lines
\citep[][require only that $\ewoii<5$\AA]{D+99}.  In particular,
H$\alpha$ is a more sensitive indicator of star formation than [OII],
and its inclusion in the selection criteria is important
\citep[e.g.][]{A2390,Q+04,BlakeEA}.  However, in this work
we will show that the properties of e(a) galaxies with only weak
emission lines, $\ewha<10$\AA\ and $\ewoii<10$\AA, are similar to
k+a galaxies (i.e. those without any emission).  We will therefore
find it convenient to exclude these from the e(a) sample, and present
them as a third class of galaxy.  Not wishing to complicate
the nomenclature further, we will simply refer to them as k+a
galaxies with weak emission.  

From this catalogue, we selected galaxies with $z>0.05$ and surface brightness
$z^\prime<21.25$ mag~arcsec$^{-2}$ for infrared observations.  The
selection was mostly random, though preference was given to k+a
galaxies over the more common e(a) galaxies.  Furthermore, we excluded
a few galaxies that showed emission line ratios clearly indicative of
an active nucleus.  For weak emission lines, the strong underlying
absorption that is characteristic of e(a) galaxies makes it difficult
to measure robust line ratios and, therefore, we cannot exclude the
possibility that some of these galaxies have a contribution to their
emission from an active nucleus.
The target list of 128 k+a galaxies (including those with weak emission) is
given in Table~\ref{tab-ka}.  This gives the positions (columns 1 and
2), redshift (column 3) and r magnitude (column 4) from the SDSS
catalogue.  The remaining columns list derived quantities that are
described later in the paper, as appropriate.
Table~\ref{tab-ea} lists the same quantities for the 94 e(a) galaxies
in the sample.
This is smaller than the k+a sample
because we preferentially observed k+a galaxies; in a
magnitude--limited sample, e(a) galaxies are at least twice as common as k+a
galaxies \citep{tomo-EA1}.

As a comparison sample, we take galaxies from the second data
release of the SDSS \citep{DR2}; 99756 of these have infrared
magnitudes that are
available from the 2MASS
\citep{2MASS} catalogue, which is complete to
$K_s=13.9$. In particular, we will use the $u$, $g$, and $r$ magnitudes
from the SDSS catalogue.  The $(u-g)$ colour is particularly important
as it has been shown to effectively divide galaxies into
two distinct populations, based on their star formation rate
\citep[e.g.][]{Strateva01_short,Baldry03}. 
These data will be compared with our k+a and e(a) galaxy
samples in relatively small redshift bins, so differences in the overall
redshift distribution are not important.

\subsection{Near infrared Observations}
Observations were made with the UFTI near-infrared imaging spectrograph
on UKIRT over two semesters.  In semester 02B we observed in classical
mode over Aug 23-30.  Weather conditions were mixed, but we obtained H
and K-band images of 144 galaxies in 4 clear nights.  In semester 03A
we were awarded an additional 33 hours of queue-scheduled observing
time to complete the observations, and another 100 galaxies were imaged
in good conditions over that semester.

The $K-$ band integrations were 600s long, dithered in a 9-point
pattern.  The exposures in $H$ were shorter (75s), and dithered in
a 5-point pattern.  Dark frames were obtained 2--3 times per night.
Data reduction was performed using the standard UKIRT pipeline
reduction tool, which aligns the dithered images and subtracts a sky
frame made from the data.
\subsection{Measurements}
\subsubsection{Photometry}\label{sec-phot}
\begin{figure}
\leavevmode \epsfysize=8cm \epsfbox{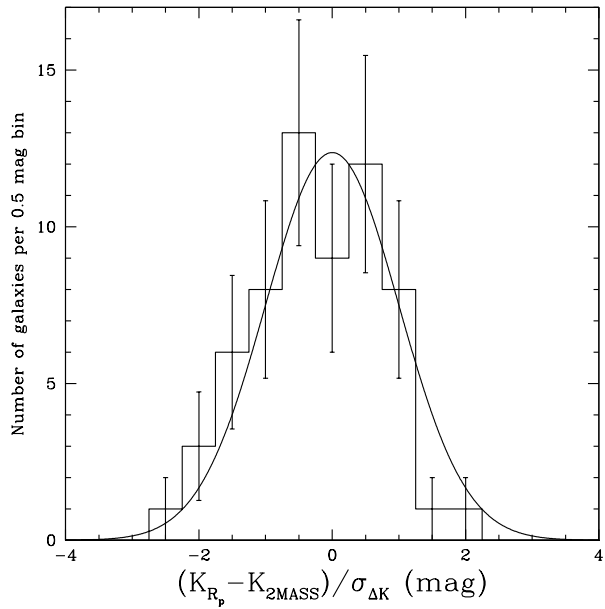} 
\caption{The histogram shows the distribution of the difference between
  our magnitudes obtained within twice the R-band petrosian radius and the Kron elliptical
magnitudes measured by 2MASS, normalised by the quadrature sum of the uncertainties on the two
magnitudes ($\sigma_{\Delta K}$).  The {\it solid line} shows a
Gaussian distribution with unit standard deviation for comparison.
This shows that our $K$ magnitudes are consistent with those of 2MASS,
within the uncertainties.
\label{fig-checkmags2}}
\end{figure}
Photometry was performed with the {\sc IRAF} task {\it qphot} by
computing the flux within a circular aperture and subtracting the sky
flux measured in an annulus outside this aperture.  
%In some cases more than one galaxy lies
%within the photometric aperture; we do not deblend the photometry, but
%flag these cases as overestimates of the luminosity.
To provide the best match with the SDSS photometry, we compute the flux
within an aperture that is twice as large as the R-band Petrosian
radius.  This magnitude,
corrected for Galactic extinction using the dust maps of \citet{SFD},
is then directly comparable to the magnitudes used to compute
Petrosian colours in the optical SDSS bands.

Standard stars were observed throughout both runs to calibrate the
photometry; however, non-photometric conditions, particularly during
the classically--scheduled first run, mean this calibration is unreliable.
To establish the zeropoint with better
precision, we compare the aperture photometry within a 5\arcsec\ aperture with the equivalent aperture
photometry from the 2MASS, where available.  Using this comparison, we
trace zeropoint changes as a function of time
through each run, and this way are able to calibrate the data to within
$\sim 0.08$ mag; this zeropoint uncertainty always dominates our
statistical uncertainty.  All our observations were made near zenith and no
airmass corrections were applied.  We do not correct our magnitudes for
Galactic extinction, as this correction is typically $\lesssim0.01$ mag in the $K-$band.

In Figure~\ref{fig-checkmags2} we show the distribution of the difference between
our total (Petrosian) magnitudes and the Kron elliptical
magnitudes measured by 2MASS, for 56 galaxies.  This difference is
normalised by the quadrature sum of the uncertainties in the two
magnitudes ($\sigma_{\Delta K}$), so the distribution should be a Gaussian with standard
deviation of unity if the magnitudes are consistent within the errors.
We include the systematic zeropoint uncertainty of 0.08 magnitudes on
our UKIRT observations.
The solid line shows this Gaussian curve for comparison; the data are
fully consistent with this distribution, which shows
that our fixed-metric apertures can be directly compared with the 2MASS
Kron elliptical magnitudes.  In absolute terms, the 1-$\sigma$ standard
deviation of the difference between the two magnitudes is 0.15 mag.
The measured magnitudes are given in column 5 of Tables~\ref{tab-ka}
and \ref{tab-ea}.

Luminosities are computed assuming a $\Lambda$CDM cosmology with the
spectroscopic redshift from SDSS.  No k-correction or evolution
correction is applied to the luminosities or colours presented in this
paper.  This is because these corrections are model--dependent, and by
their nature the galaxies in this sample may have unusual star
formation histories which make the usual models inapplicable.  Instead,
we will compare samples within the same, relatively narrow, redshift ranges.

\subsubsection{Morphology measurements}
Galaxy morphologies are determined by fitting a two-dimensional
parametric model to the surface brightness distribution, using the {\sc
  GIM2D} software \citep{GIM2D}.  The model
consists of a bulge and disk component, and is completely described by
twelve parameters.  From these
fits, we derive the ratio of the bulge luminosity to the total
luminosity, $B/T$.  Since the small UFTI field of view does not
generally contain enough stars to obtain a reliable point-spread
function (PSF) for each image, we use the PSF from the standard star
observed most closely in time (generally within 2 hours).  
We find no significant correlation between the measured B/T and the
seeing or redshift, from which we conclude that uncertainties in the
PSF do not significantly influence our B/T measurements.  The measured
B/T values are given in column 6 of Tables~\ref{tab-ka} and
\ref{tab-ea}.  The formal uncertainties in B/T computed by {\sc gim2d} are
quite small, $<0.1$; however, the true uncertainty is
likely to be dominated by
systematic errors (such as PSF fitting and the assumption of a
$r^{1/4}$ law for the bulge) on the order of $\sim 0.1$--$0.2$.  We
also note that the $B/T$ will be meaningless in irregular systems of
close pairs or merging galaxies.  

In Appendix~\ref{sec-morph} we show images and the corresponding surface
brightness fits for most\footnote{We show all of the k+a galaxies, but
  for the sake of brevity just a random subsample of
  48 of the 57 k+a galaxies with weak emission, and 60 of the 94 e(a)
  galaxies.} of the galaxies in our sample, grouped by B/T and emission line strength
(Figures~\ref{fig-EAtrue_disk}--\ref{fig-EAem_bulge}).  For each galaxy
we show the central 
12\arcsec of the original $K-$band image, as well as the {\sc GIM2D} best-fit model (with logarithmically spaced
contours), and the residuals
between the two.  A visual comparison shows that the GIM2D
B/T ratio provides a reasonable morphological classification; the low
B/T galaxies mostly have obvious disks, while the high B/T galaxies are
generally spheroidal.  The method clearly does not deal well with
galaxies that are distorted or have twisted isophotes; however, most of
the galaxies in our sample have a normal appearance.

We also visually inspect each image for obvious signs of interactions, and we
mark any galaxy that shows the strong morphological distortion
typical of tidal effects, or a close companion and some
indication for tidal distortion, as a possible interacting system.  Our
images are not deep enough to search for low surface--brightness tidal
features, however.  Furthermore, the completeness of our list of
interacting galaxies is uncertain without doing a careful analysis of
how surface brightness dimming with redshift and differences in galaxy surface
brightness profile alter the detection of these features
\citep[e.g.][]{YZZ+}.   We therefore do not present a quantitative analysis of
these potentially interacting galaxies in this paper.

\subsection{Stellar population modelling}\label{sec-models}
\begin{figure}
\leavevmode \epsfysize=8cm \epsfbox{Fig2.epsi} 
\caption{A colour--colour diagram of normal ($r$-selected) galaxies in the SDSS,
  compared with population synthesis models from \citet{BC03}.  
  These represent our fiducial models, which are intended to be representative of
  typical galaxies in the local Universe.  The
  {\it small dots} are bright ($r<16.4$) SDSS galaxies at
  $z=0.06\pm0.01$ without strong
  emission lines ($\ewha<2$\AA\ and $\ewoii<2$\AA), while the small,
  {\it open circles} are galaxies at the same redshift and magnitude,
  but with $\ewha>40$\AA\ and $\ewoii>20$\AA.  The {\it filled circle}
  shows a single--burst model, with no dust extinction, and a
  \citet{Sp} initial mass function, after 13.7 Gyr of passive
  evolution.  The smaller arrow (pointing to redder colours) shows the
  effect of increasing the dust optical depth from 0 to $\tau_v=1$,
  with 30 per cent of the dust arising from the ambient ISM.  The
  longer arrow (toward bluer colours) shows the effect of decreasing
  the metallicity to 40 per cent solar.  The {\it large star} shows a model with constant star
  formation rate and dust extinction with a total optical depth
  $\tau_v=1$.  The redward arrow shows
  the effect of increasing the dust extinction to $\tau_v=2$.  The
  near-vertical arrow shows the effect of
  reducing the metallicity to 40 per cent solar.  Finally, the {\it
    solid line} shows the evolutionary track of a galaxy with an
  exponentially declining SFR, with timescale 4 Gyr and $\tau_v=1$.
\label{fig-calib}}
\end{figure}
To help interpret our observations we use the \citet{BC03} population
synthesis models.  For our fiducial model we assume a \citet{Sp} initial mass function with
solar metallicity.  We first verify that reasonable models are able to reproduce the colours of normal 
galaxies by comparing with the $r-$selected SDSS data in
Figure~\ref{fig-calib}.  The simplest case is to compare a
single--burst, dust--free model, 
evolved for a Hubble time (13.7 Gyr), with bright (i.e. high
signal--to--noise) SDSS galaxies which have no detectable \oii\ or
H$\alpha$ emission lines.  The model prediction is shown as the large,
filled circle, and the data are the small points;  
the agreement is good, and improved if we include a range of
abundances for the early type population:
a model with lower metal abundance, 40 per cent of solar,
yields bluer colours as indicated by the longer arrow.  Dust extinction
changes the colours in the other direction, as shown by the shorter
arrow.  This estimate is based on the two-component model of
\citet{CF00}, with a total optical depth of $\tau_v=1$, of which 30 per
cent arises from the ambient ISM and the remainder is due to molecular
clouds.  Thus younger, more massive stars are more heavily extincted
than the older stellar population.

To compare the
model predictions for normal, star--forming galaxies we compute a model
with a constant star formation rate over a Hubble time, and include
dust extinction of $\tau_v=1$, again assuming 30 per cent of the
extinction arises from the ambient interstellar medium.   This type of model is known
to be a reasonable representation of local, star--forming galaxies
\citep[e.g.][]{Brinchmann03}.  The model is presented as the large
star in Figure~\ref{fig-calib}, and is compared with SDSS galaxies with $\ewha>40$\AA\ and
$\ewoii>20$\AA\ (open circles).  Again the agreement is good, and the
vectors show the effect of increasing the extinction to $\tau_v=2$ or
decreasing the metallicity to 40 per cent of solar.
The solid line shows the evolutionary track of a galaxy with a SFR that
declines exponentially, with a timescale of 4 Gyr, and $\tau_v=1$; it
nicely traces the locus of normal emission-line galaxies in the SDSS as
it evolves over a Hubble time (from blue to red colours).
  
These models illustrate the usefulness of the $(u-g)$, $(r-k)$ colour combination.  The
$(u-g)$ colour is sensitive to population age (it differs by almost 1
magnitude between the two models), but is relatively weakly sensitive to
metallicity and, especially, dust effects.  On the other hand, the
$(r-k)$ colours are approximately equally sensitive to age, metallicity
and dust.  

\section{Results}\label{sec-results}
\subsection{Galaxy morphologies}
\begin{figure}
\leavevmode \epsfxsize=8cm \epsfbox{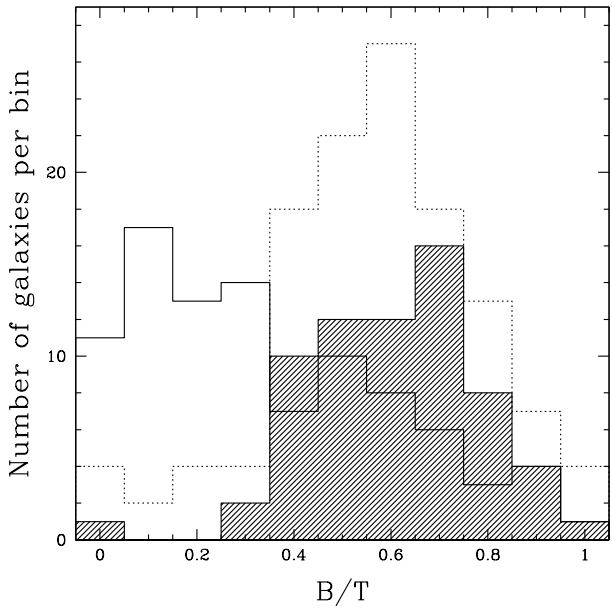}
\caption{The distribution of B/T for the k+a galaxies ({\it shaded
    histogram}) and e(a) galaxies ({\it solid line}).  The {\it dotted
    line} shows the result of adding galaxies with only weak
  emission lines ($\ewha<10$\AA, $\ewoii<10$\AA) to the k+a sample.
  The distributions of k+a and e(a) galaxies are distinctly different, in the sense that k+a
  galaxies are primarily bulge--dominated, while e(a) galaxies are
  mostly disk--dominated.
\label{fig-btdist}}
\end{figure}
Figure~\ref{fig-btdist} shows the distribution of B/T for the k+a and
e(a) galaxies in the sample.  
The two types of galaxies have distinctly different morphologies
(significant at the $>3\sigma$ level): while
almost all the k+a galaxies are bulge--dominated, the opposite is true
for e(a) galaxies. Similar results have been found by \citep{Q+04},
based on measurements of the Sersic profiles rather than B/T ratios.  Including galaxies with weak emission
($\ewha<10$\AA, $\ewoii<10$\AA) in the k+a sample does not change this
result.  Thus, galaxy morphology is correlated more strongly 
with the instantaneous star formation rate (i.e. emission lines) than with recent
star formation over the last $\sim 1$ Gyr as represented by the H$\delta$ absorption line.  

A closer examination of the disk--dominated k+a galaxies in
Figure~\ref{fig-EAtrue_disk} shows that most have a
very smooth disk component, with little or no evidence for spiral
structure in the model residuals.  
This contrasts with the e(a) disk--dominated galaxies shown in
Figure~\ref{fig-EAem_disk}, which show morphologies more akin to
typical spiral galaxies.  
The spheroid--dominated k+a galaxies appear mostly normal, with few signs
of disturbance or interactions, although our data are not deep enough
to identify the subtle features expected from recent merger activity \citep{YZZ+}.

\subsection{Environment}
\begin{figure}
\leavevmode \epsfxsize=8cm \epsfbox{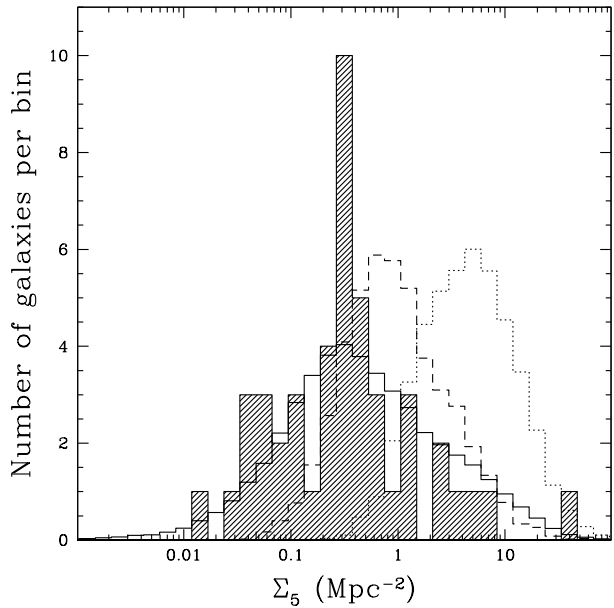}
\caption{The distribution of projected local surface density $\Sigma_5$
  for the k+a population at $0.05<z<0.1$ ({\it shaded histogram}, including galaxies with
  weak emission lines) compared with the
  $r-$selected SDSS subsample in the same redshift range {\it (solid line)}.  The remaining
  lines  show the SDSS distribution restricted to
  galaxies in clusters with $\sigma>400$ km~s$^{-1}$: the {\it dashed
    line} represents galaxies within $1<R/R_{\rm vir}<3$ and the {\it
    dotted line} represents galaxies with $R<R_{\rm vir}$, where
  $R_{\rm vir}$ is the cluster virial radius.  The k+a galaxies are
  found in typical environments, and not preferentially in or near clusters.
\label{fig-sigdist}}
\end{figure}
We can characterize the galaxy environment by the number of
neighbouring galaxies \citep[e.g.][]{Dressler}.  Specifically, we measure $\Sigma_5$, the
density derived from the projected distance to the
fifth--nearest neighbour brighter than $M_r=-20$, as in \citet{BB04}.
This gives a useful estimate of the local density for galaxies at
$z\lesssim 0.1$; at higher redshifts the spectroscopic limit is
brighter than $M_r=-20$, and $\Sigma_5$ is therefore an underestimate.
We therefore restrict our sample to galaxies in the redshift
range $0.05<z<0.1$.

In Figure~\ref{fig-sigdist} we show the distribution of $\Sigma_5$ for
the 46 k+a galaxies (including those with weak emission) in the redshift
range $0.05<z<0.1$, compared with the distribution for all galaxies in
that redshift range from the SDSS data (including those without
available infrared magnitudes from 2MASS).  The two distributions are
statistically consistent with being drawn from the same population.
Thus, k+a galaxies at $0.05<z<0.1$ are not more likely to be found in clusters than the
typical $r-$selected galaxy in the SDSS.  For comparison,
we also show the distribution of $\Sigma_5$ for the subset of the SDSS
galaxies found in clusters with velocity
dispersion $\sigma>400$ km~s$^{-1}$.  The clusters are selected from
the C4 catalogue of \citet{C4}, and we divide the galaxies
into two populations based on their projected position relative to the
cluster virial radius, $R_{\rm vir}$.  In the virialised regions of
clusters, $r<R_{\rm vir}$, densities are typically
$\Sigma_5\gtrsim2$~Mpc$^{-2}$, much larger than the densities of all
but a few of our k+a galaxies.  Even when considering only
galaxies well outside the virial radius of these clusters, $1<R/R_{\rm
  vir}<3$, local
densities are typically $\sim3$ times larger than average, and inconsistent with
the k+a galaxy population.  We therefore confirm the conclusions of others
\citep{Z+96,Q+04,BlakeEA}, that nearby k+a galaxies do not preferentially
reside within clusters and rich groups; moreover, they are not more likely
to be found in these dense environments than typical,
$r-$selected galaxies.

\begin{figure}
\leavevmode \epsfxsize=8cm \epsfbox{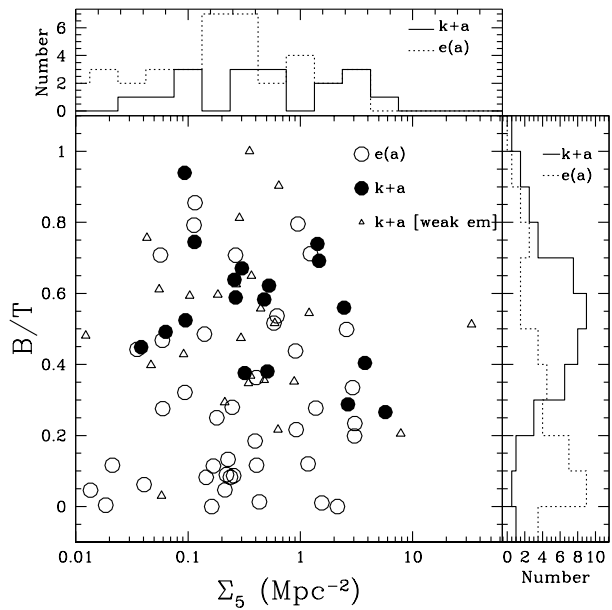}
\caption{The morphologies of galaxies with $0.05<z<0.1$ are shown as a
  function of their local galaxy density.  The {\it open circles} show
  the e(a) galaxies, while the {\it solid circles} represent the k+a
  galaxies.  {\it Triangles} are k+a galaxies with a small amount of
  emission, $\ewha<10$\AA\ and $\ewoii<10$\AA.  The histograms on the top and left of the figure show the
  collapsed distributions of $\Sigma_5$ and B/T for the k+a galaxies
  (including those with weak emission) and e(a)
  galaxies, as indicated.  Although k+a and e(a) galaxies are found in
  similar environments, their morphologies are distinctly different.
\label{fig-plotenv}}
\end{figure}

Although k+a galaxies are found in both high and low density regions, it is
interesting to investigate whether those few that are found in clusters
differ in any significant way from the rest of the population.  In
Figure~\ref{fig-plotenv} we show the B/T 
ratio of the galaxy sample with $0.05<z<0.1$, as a function of $\Sigma_5$.
The small sample means that it is difficult to robustly identify any trends with
environment, and a Spearman's rank correlation test
does not find a significant correlation.  However, we note that in the
densest regions, $\Sigma_5>2$~Mpc$^{-2}$, we do not find any galaxies
with B/T$>0.6$.  Although this difference is not statistically
significant, neither can we rule out the possibility that k+a galaxies
in clusters are mostly disk--dominated \citep[e.g.][]{TranEA,Caldwell}, and are a distinct phenomenon
from the bulge--dominated galaxies in the field.

The e(a) galaxies, while morphologically quite distinct from the k+a
galaxies,  inhabit similar environments.  The distributions and sample
means of $\Sigma_5$ for the two populations are statistically consistent with being drawn from the same
population, as determined with a Kolmogorov--Smirnov
test and a Students t-test, respectively.  Thus the difference between
these two types of galaxy does not appear to be related to environment.

\subsection{Optical and infrared magnitudes}\label{sec-colours}
\begin{figure}
\leavevmode \epsfxsize=8cm \epsfbox{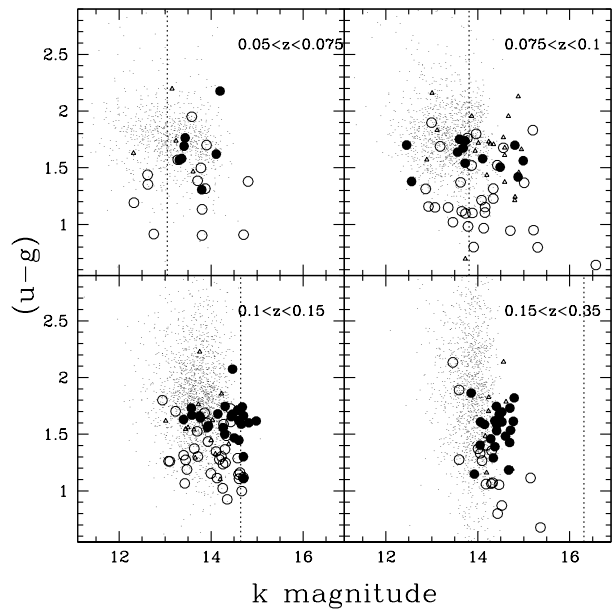}
\caption{The observed (u-g) colours as a function of $k$ magnitude for
  the k+a galaxies ({\it filled circles}) and e(a) galaxies ({\it open
    circles}).  {\it Triangles} represent k+a galaxies with weak
  emission lines. The {\it small dots} are normal SDSS galaxies for which
  $K-$magnitudes are obtained from 2MASS, and which show no H$\alpha$
  emission ($\ewha<4$\AA). We only show a random 10 per cent of the
  galaxies for clarity.  The {\it dotted lines} show the
  characteristic luminosity $M^\ast=-24.2$ from
  \citet{Cole-2mass_short}, neglecting small k- and
  evolution-corrections.  Our magnitude selection effect means the
  nearby galaxies are mostly faint, while the highest redshift sample
  includes only the most luminous galaxies.
\label{fig-ugK}}
\end{figure}
\begin{figure}
\leavevmode 
\epsfxsize=8cm\epsfbox{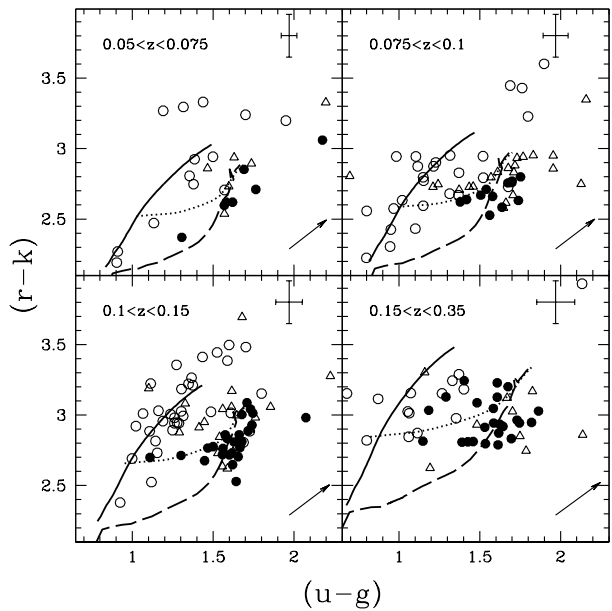}
\caption{
The $(u-g)$--$(r-k)$ colour--colour diagram in
  four redshift bins.  
The {\it open circles} represent the
  e(a) galaxies, and the {\it filled circles} are the k+a galaxies;
  {\it triangles} represent k+a galaxies with some weak emission.  The sample error bars in the top right of each panel
  show the median 1$\sigma$ uncertainties.   The
  lines show three different models from \citet{BC03}, as described in
  the text.  The arrow in the bottom-right of each
  panel shows the change in colour corresponding to dust
  extinction of $\tau_v=1$.  The {\it solid
    line} shows a galaxy with $\tau_v=1$ and an exponentially declining star formation
  rate, with $\tau=4$ Gyr; colours become redder with time.  The {\it dotted line} shows the effect
  of truncating star formation in a galaxy that was forming stars at a
  constant rate for 13.7 Gyr (i.e. a Hubble time).  The {\it dashed line}  
  shows the
  evolution of a galaxy with an old stellar population after the
  addition of a burst accounting for $15$ per cent of its stellar
  mass.  Evolution is from left to right, and both models assume no
  dust extinction.   Further
  details on the models are given in the text.  The k+a and e(a)
  galaxies are distinctly separate populations in this plane, and
  generally inconsistent with the truncated--star formation model
  (dashed line).  
\label{fig-ugri}}
\end{figure}
The photometric data are presented as a colour--magnitude diagram in
Figure~\ref{fig-ugK}, where we compare the optical (u-g) colours of our sample with
the UKIRT k magnitudes, in four redshift bins.  
For comparison we also show as small dots those SDSS galaxies with
$\ewha<4$\AA, for which a K-band magnitude is available from the
shallower 2MASS.  Our $r-$selected e(a) and k+a galaxies have $k$
magnitudes as faint as $k\sim15$, well below the 2MASS limit. Because
the sample is magnitude--limited, the
luminosity distribution is strongly redshift--dependent.
We show the
characteristic magnitude $K^\ast$ from \citet{Cole-2mass_short},
$M=-24.2$, at the midpoint of each redshift bin, neglecting k- and
evolution-corrections, which are small for normal galaxies (generally $<0.2$ mag).  
In our highest redshift bin ($z>0.15$), which contains $\sim 30$ per
cent of the sample, the galaxies are very luminous, $\sim 1.5$ mag
brighter than $L^\ast$.  On the other hand, at the lowest redshifts,
$0.05<z<0.075$, all seven of the k+a galaxies 
observed are fainter than $K^\ast$, by up to $\sim 1$ mag.  
Note that
this is consistent with the results of
\citet{PoggEA03} who find that the k+a population in Coma consists mostly
of faint galaxies.  However, the lack of brighter galaxies in our
small low-redshift sample is only inconsistent with the normal, passive galaxy
distribution at the 2$\sigma$ level.
Most of the k+a
and e(a) galaxies are bluer in (u-g) than the red sequence of typical SDSS galaxies lacking
H$\alpha$ emission.  This is not surprising,
since the e(a) galaxies have emission lines indicative of ongoing star
formation, and the strong H$\delta$ absorption of the k+a galaxies
likely indicates recent star formation activity.

\section{Interpretation and modelling}\label{sec-interp}
\subsection{Optical and infrared colours}
We show the
optical--infrared colour $(r-k$) as a function of $(u-g)$
in Figure~\ref{fig-ugri}.
The solid line in the figure is the normal spiral evolution track
introduced in Figure~\ref{fig-calib}; it
 is evident that most of the
e(a) galaxies lie along this locus.  The tightness of the colour
relation is surprising, and may break down at the highest redshifts,
$z>0.15$, where unfortunately our sample of e(a) galaxies is small.  It
is also noteworthy that many e(a) galaxies lie redward of the end of
the normal spiral track (and redder than normal star--forming galaxies;
recall Figure~\ref{fig-calib}), in the direction expected if they are
more than normally reddened by dust.
 
In contrast, the k+a galaxies  have colours
that are distinct from those of most normal
galaxies in general, and e(a) galaxies in particular.   In particular, they have $(u-g)$ colours
that are intermediate between the passive and star--forming population,
as we saw in Figure~\ref{fig-ugK}, while their $(r-k)$ colours
are similar to those of the normal star--forming population.  
Note also that galaxies with only weak emission (the triangles in the
figure) have colours most similar to the k+a population, rather than
the e(a) population or something in between the two.

To interpret these colour distributions, we consider two distinct 
variations on the normal galaxy models introduced in \S~\ref{sec-models}.  It is not our goal to explore the full
range of parameter space; degeneracies with dust, initial mass
function, and metallicity make it difficult to make robust statements
about the star formation history.  However, we will contrast two very
different models as an illustration of the
two broad paths generally proposed to generate a k+a spectrum:
either truncating star formation in a normal disk galaxy, or introducing a
short burst of star formation on an older population.  We model the
first case (``truncation'') by taking a galaxy with a constant star formation rate and
truncating star formation after a Hubble time (13.7 Gyr, at which point the integrated
colours are similar to those of normal spiral galaxies); this model is shown as
the dotted line in Figure~\ref{fig-ugri}.  We assume no dust extinction
in this model, based on the assumption that even if the initial galaxy
has substantial dust, the termination of star formation (whether by
merger--induced starbursts or gas stripping, for example) would be accompanied by a
depletion of dust as well.
At the time when star formation ceases, the model
quickly becomes redder in the star--formation sensitive $(u-g)$ colour; however the 
$(r-k)$ colour remains approximately constant.  Approximately 300 Myr
later, $(r-k)$ also begins to redden.  Furthermore, the
$(u-g)$ colour
evolution slows down: it takes $\sim 1.4$ Gyr for $(u-g)$ to redden by only another
0.2 magnitudes.  The observed population of k+a galaxies (and a few
e[a] galaxies) have $(u-g)$ colours that are similar to this transition point, a few hundred Myr after
the truncation of star formation; however, the $(r-k)$
colours of many are too blue.  This is a generic problem of
the assumption that the galaxy starts
from a colour that is typical of normal spiral galaxies, since $(r-k)$
can only get redder after the truncation of star formation, and including dust extinction in
the model makes the situation worse.  The only way to match the
$(r-k)$ colours with this kind of model is to start with a bluer
galaxy, initially, than a galaxy with a constant star formation rate.
One way to do this would be with a current (or very recent) star
formation rate that exceeds its past average; this becomes a weaker
version of the ``starburst'' model considered below.  
We can therefore conclude
that the $(r-k)$ colours of many k+a galaxies are inconsistent with
the assumption that they form via the truncation of star formation in normal spiral galaxies.

\begin{figure*}
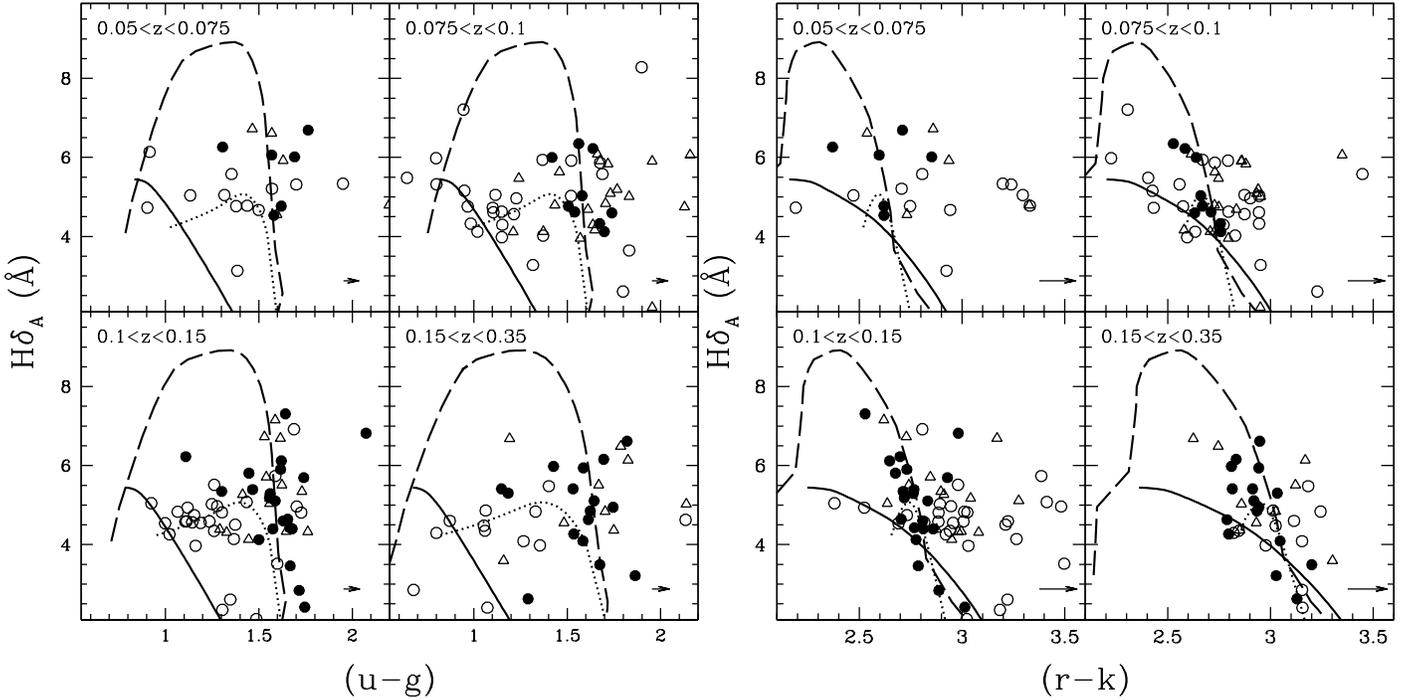

\leavevmode \epsfxsize=9.25cm \epsfbox{Fig8a.epsi}\epsfxsize=9.25cm\epsfbox{Fig8b.epsi}
\caption{The H$\delta$ line strength, defined as in \citet{WO97}, is shown as a function of $(u-g)$
  colour ({\it left}) and $(r-k)$ colour ({\it right}).  The {\it open
    symbols} are the e(a) galaxies, and the {\it solid circles} are k+a
  galaxies.  {\it Triangles} are k+a galaxies with a small amount of
  nebular emission.  The lines show the
  same models as in Figure~\ref{fig-ugri}, and the arrow in the lower
  right corner shows the effect of increasing the dust extinction by 1
  magnitude in a galaxy with a constant star formation rate.  Note that
  the $H\delta_A$ predictions for the normal spiral model ({\it solid
    line}) are upper limits, as emission--filling will reduce the
  equivalent width.  The k+a galaxies show good agreement with the
  burst model (dashed line), while many of the e(a) galaxies are particularly
  noteworthy for their very red $(r-k)$ colours, suggestive of strong
  dust obscuration.
\label{fig-ugHd}}
\end{figure*}
To investigate the alternative case, we start with an evolved (13.7 Gyr),
dust--free, single
burst population, which has colours typical of the red locus of
SDSS galaxies.  We then add a burst comprising 15 per cent of the total
mass of stars formed and evolve the model for another 2 Gyr; this is
shown as the dashed line (which starts from the end of the burst and
evolves redward).  In this case, the colour
evolution provides an excellent match to the average colours of most k+a
galaxies.  We cannot draw strong conclusions
regarding the strength or age of the burst based on these data, however.  Even for this
particular choice of parameters (i.e. metallicity, dust, initial mass function) the scatter
in the colours admits any burst strength $>5$ per cent. 

It is evident is that the e(a) population in general does not lie along
either model track: their colours resemble normal spiral galaxies more
than they do an early phase of the k+a galaxies.  This is at least true
for all but the highest redshift galaxies in our sample.  At $z>0.15$
the e(a) and k+a populations may be more consistent with a single
colour sequence, 
approximately following the truncated spiral galaxy model.  However,
there are few data here, and they span a significant
redshift range which will increase the scatter in the colours.  If the
difference relative to the lower redshift bins is real, it may be
an evolutionary effect, or it may be a luminosity effect, as the
highest redshift k+a and e(a) galaxies are up to $\sim 3$ magnitudes
more luminous (and therefore likely more metal-rich) than their low--redshift counterparts.

\subsection{Colours and H$\delta$ line strengths}
In Figure~\ref{fig-ugHd} we show the observed and
model dependence of H$\delta$ line strength on $(u-g)$ and $(r-k)$ colour.  
The models are the same ones presented in Figure~\ref{fig-ugri}.  In
order to make the comparison between the models and data fair, the line
strengths for both
are remeasured following the definition of \citet{WO97}. These measurements
differ from the more sophisticated measurements of  \citet{tomo-EA1}, which are based on a model
Gaussian fit to the data; to distinguish them we adopt the notation of
\citet{WO97} and label them $H\delta_A$.  In general the two measurements are
comparable, but there are a few cases where the \citet{WO97} definition
yields a significantly lower value.  The original \citet{tomo-EA1}
measurements are given in column 7 of Tables~\ref{tab-ka} and
\ref{tab-ea}, while our remeasured H$\delta_A$ are listed in column 8.
For completeness, the rest-frame equivalent widths of [O{\sc ii}] and
H$\alpha$ from the SDSS pipeline are also given in those tables as
columns 9 and 10, respectively.

As has been noted many
times before \citep[e.g.][]{CS87,P+99,PSG}, truncating star formation
in a normal spiral galaxy
(i.e. one that has been forming stars at a constant rate for many Gyr)
does not produce strong enough $\ewhd$ to match the strongest--lined k+a
galaxies.  
However, the relatively narrow colour range of most of the k+a galaxy population is in
remarkably good agreement with the 15 per cent burst model, for the
range of observed $\ewhd$ strengths.  The galaxies with the strongest
H$\delta$ lines are the most convincing: only the burst model can match
the relatively red $(u-g)$ colours and the blue $(r-k)$ colours,
simultaneously with the strength of the absorption. Again the burst strength is not
well constrained, and any burst making up at least 5 per cent of the
stellar mass is in agreement with most of the data.   

On
the other hand, the e(a) galaxies show a large amount of scatter, and
do not appear to form a single sequence.  Some of this scatter will be
caused by emission--filling of H$\delta$ absorption, which is not
accounted for, and means that $H\delta_A$ is an underestimate of the
underlying absorption.  Although in many cases the $(u-g)$ colours and
H$\delta$ line strengths are approximately consistent with the
truncated spiral model, the $(r-k)$ colours are much too red for this
to be the correct model (also, of course, the presence of emission
lines in the spectrum means a completely truncated model cannot be
physically correct).
In fact, the $(r-k)$ colours
of the reddest galaxies are not matched by any of the models presented here;
since this colour is much more sensitive to dust extinction than $(u-g)$,
one possible interpretation is that e(a) galaxies are more heavily
extincted than normal spirals.  For most of the population an extra
$\sim 1$ magnitude of dust extinction would be consistent with their position on the colour--colour diagram of
Figure~\ref{fig-ugri}.  Conversely, the lack of scatter in
$(r-k)$ colour for the k+a population means that dust is unlikely to
play a strong role in the spectral properties of these galaxies.

\subsection{Stellar luminosities and masses}
\begin{figure}
\leavevmode \epsfxsize=8cm \epsfbox{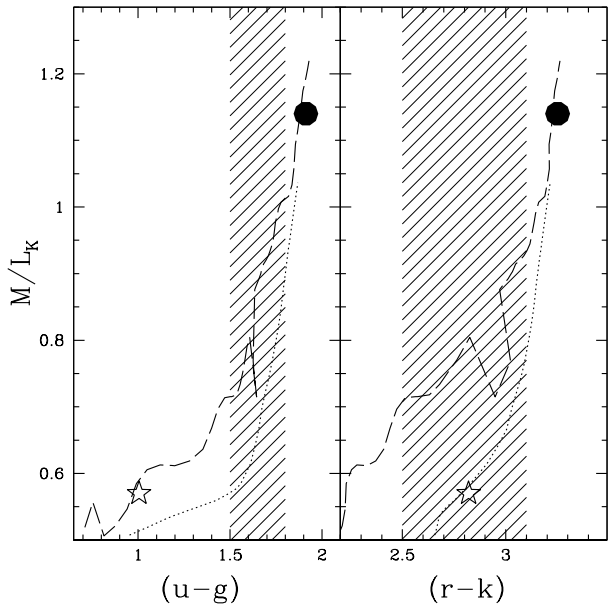}
\caption{The $M/L_K$ ratio for four \citet{BC03} models at $z=0.12$, as a function
  of $(u-g)$ colour (left) or $(r-k)$ colour (right).  The {\it solid
    circle} and {\it large star} represent the old and young fiducial
  models, as presented in Figure~\ref{fig-calib}.  
The {\it dotted line} shows the 
  truncated-disk model, where star formation is abruptly stopped after 13.7
  Gyr and there is no dust.  The {\it dashed line}
  shows the track for a 15 per cent burst of star formation on top of
  an old stellar population, also without dust.  The {\it shaded region} shows approximately the colour
  range of observed k+a galaxies at $0.1<z<0.15$, from
  Figure~\ref{fig-ugri}.  If typical k+a galaxies result from a 15 per
  cent burst superposed on an old stellar population, this indicates
  they will have $M/L_K\sim 0.8\pm0.1$.  
\label{fig-ML}}
\end{figure}
The $K-$band luminosity is a good tracer of stellar mass, and is less
sensitive to the recent star formation history than optical
luminosities.  However, the $M/L_K$ ratio can still vary by a factor
$\sim 2$ depending on population age;
in Figure~\ref{fig-ML} we show how $M/L_K$ 
depends on $(u-g)$ and $(r-k)$ colour for the models considered
in this paper.  The fiducial model of an old galaxy population,
presented in Figure~\ref{fig-calib}, has $M/L_K\sim1.15$, while the
constant star--formation model (with $\tau_v=1$ dust extinction)
has $M/L_K\sim 0.55$.  Thus, the bluest normal galaxies are typically a factor $\sim
2$ less massive then red galaxies of the same $K-$luminosity.  Our best
interpretation for most of the e(a) population is that they are normal
(though possibly dusty) spiral
galaxies, which means they will also have $M/L_K\sim0.55$.
We also show
the models of truncated star formation in a normal disk galaxy, and of a 15 per
cent burst superposed on an old stellar population.  Recall that this
latter model provides a good match
to the colours, H$\delta$ line strengths, and morphologies of the k+a
population, but the burst strength is poorly constrained only to be $>5$ per
cent by mass.
This  model
predicts $M/L_K\sim 0.8\pm0.1$ at the colours typical of the k+a
galaxies at $z\sim 0.12$, not
too dissimilar from  the value one would obtain by interpolating
between the evolved and star--forming models, as a function of $(u-g)$.
Thus, if the model we have chosen to represent the colours of k+a
galaxies is correct,
then they have $\sim 75$ per cent of the mass of normal,
passive galaxies at the same K-band luminosity.  The typical $M/L_K$
will be higher, $\sim 1\pm 0.1$ if the burst only makes up 5 per cent
of the final mass, and more consistent with the $M/L_K$ of normal
passive galaxies.

\section{Discussion}\label{sec-discuss}
\subsection{The connection between e(a) and k+a galaxies}
The observations and modelling results presented in this paper suggest
that most e(a) galaxies are not directly related, in an evolutionary
sense, to k+a galaxies.  This conclusion is based primarily on the fact
that, although the range of H$\delta$ absorption strengths are similar,
the colours and infrared morphologies of the two populations are quite
distinct.  
In particular, the insensitivity of $K$ luminosity to recent star formation provides
a strong morphological argument against the interpretation that k+a galaxies arise
following the end of star formation in e(a) galaxies.  From
Figure~\ref{fig-btdist} we saw that most k+a galaxies have B/T$\sim0.6$, while
most e(a) galaxies have B/T$\sim 0.1$.  If we assume that all star
formation occurs in the disk component, then a termination of star
formation means the disk will fade by only a factor $\sim 2$
(Figure~\ref{fig-ML}).  Thus the typical e(a) galaxy that started with
B/T$=0.1$ would end up with B/T$=0.18$, where the disk is still much
more prominent than for typical k+a galaxies.  Conversely, to reproduce the
typical morphology of k+a galaxies through disk fading, the progenitor
would have to have B/T$=0.42$.  Therefore this simple fading mechanism
cannot connect the two different morphological distributions, and this
conclusion is only strengthened if some
of the star formation takes place in the bulge, rather than the disk \citep[e.g.][]{N+01}.
Although this conclusion holds for about half the e(a) population, we
note that the distribution of e(a) morphologies includes a 
significant tail extending to bulge--dominated systems, which are
morphologically consistent with being progenitors of k+a galaxies.

The colours of most e(a) galaxies are sufficiently red
that it is unlikely their colours will evolve to
match those of k+a galaxies once star formation stops.  This does not
mean that none of the e(a) galaxies will  evolve into k+a types; the
bluest $\sim 10$ per cent of the population may be starburst systems
that could represent k+a progenitors.  However, most of these blue e(a)
galaxies still have disk--dominated morphologies, as shown in Figure~\ref{fig-EAem_blue}.  It is possible that
a subsequent merger event could destroy this disk, and in fact, about
half of the bluest e(a) galaxies show some suggestive evidence for interaction
with a nearby companion.  However, it is not clear if this abundance of
interactions is better correlated with galaxy colour or H$\delta$ line
strength, so we cannot conclude that the e(a) spectrum is a unique
characteristic of interacting systems destined to become k+a galaxies.
Finally, we note that if the colours of e(a) galaxies are 
the result of unusually strong dust obscuration, then more of them will
have blue intrinsic colours; as long as the dust is destroyed with the
cessation of star formation (for example, in a starburst), their colours might evolve to be
consistent with those of k+a galaxies.

Although we conclude that e(a) galaxies are not all progenitors of k+a
galaxies, we note that the argument does not work in reverse.  It is possible that
the progenitors of today's k+a population all had an e(a) spectrum in
the past.  These progenitors would have to be either a subset of the
e(a) galaxies observed today (those that are blue and disk-dominated)
or perhaps an entirely different population of very blue e(a) galaxies
that are so short-lived they are not present in our sample.
%In other words, the e(a) phase may still
%represent the primary route through which k+a galaxies are formed.
However it seems likely that selection based on H$\delta$ strength 
alone is not an efficient way to find the progenitors of k+a galaxies.

\begin{figure}
\leavevmode \epsfxsize=8cm \epsfbox{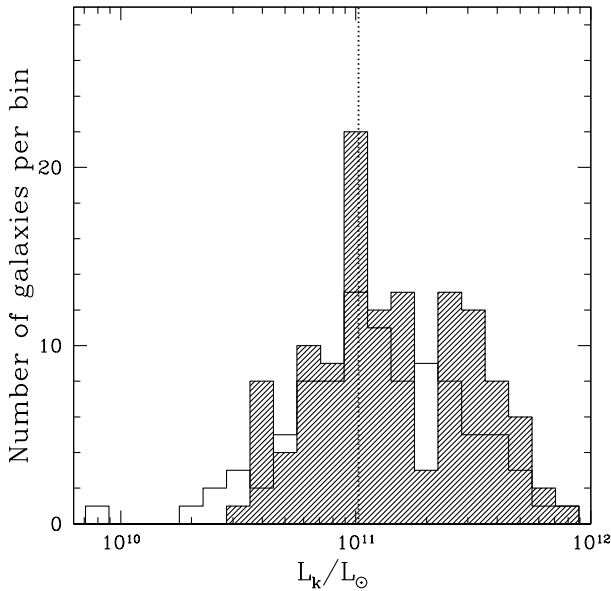}
\caption{The K luminosity distribution, in solar units, of the k+a
  galaxies ({\it shaded histogram}, including galaxies with weak
  emission lines) and the e(a) galaxies ({\it solid
    histogram}).  This is just the observed distribution of the sample
  and is not corrected for magnitude (or other) selection effects.  The two distributions are not significantly
  different, demonstrating that the observed differences in morphology and colours
  are not due to differences in luminosity.  The {\it dotted line} shows the characteristic
  luminosity, from \citet{Cole-2mass_short}, for reference.
\label{fig-Kdist}}
\end{figure}

We will consider the mass function of k+a and e(a) galaxies in a
subsequent paper; here we present just the observed
distribution of K-band luminosities, uncorrected
for selection effects, to ensure that our comparisons of the two
populations are based on samples of similar stellar masses.  In Figure~\ref{fig-Kdist} we show the
luminosity distributions
in solar units, using $K_\odot=3.33$ \citep{AAQ}, neglecting
k- and evolutionary corrections.  We include galaxies with weak
emission lines in the k+a sample.
From the mass--to--light ratio calculation computed above we see that
the mode of our k+a sample is at $M\sim 8\times10^{10} M_\odot$; for
the e(a) galaxies (which we interpret as normal spiral galaxies) the 
mode is only 25 per cent lower, at $M\sim 6\times10^{10} M_\odot$.
This only shows that when comparing the two types of galaxies we are
comparing objects of similar mass; of course the shape of the
distributions are heavily biased by our selection function.

\subsection{Comparison with previous work}
Our data are most consistent with the interpretation that local k+a galaxies
are a short--lived phase that occurs following a starburst in a
spheroidal galaxy;  a likely
cause for such a burst is a recent merger event.
Our results are generally consistent with those of \citet{N+01} and \citet{YZZ+},
who find the dynamics and morphologies of local k+a galaxies are
typical of early--type galaxies.  These authors claim the origin is
from the merger of two galaxies, at least one of which is gas--rich, based on tidal features in
some galaxies \citep{Z+96,N+01} \citep[see also][]{SwinbankEA}.  
Although we do not see
convincing evidence that a substantial fraction of our k+a galaxies are
currently undergoing mergers, our images are not deep enough to see low
surface brightness tidal features indicative of recent interactions. 

The lack of bright, k+a galaxies in dense environments is consistent
with the results of \citet{PoggEA03}, who find that k+a galaxies in
Coma are mostly much fainter than $L^\ast$.  However, our sample is too small
to claim more than this consistency, due to the general rarity of bright galaxies and dense
environments.  In distant clusters, there is evidence that
disk--dominated k+a galaxies 
may be more common \citep{TranEA}, and therefore could arise
through a different mechanism from the bright, field k+a galaxies
studied here.  Our results are not inconsistent with this
interpretation, as the few k+a galaxies we find in dense environments
do have $B/T<0.6$; however, there are too few galaxies to provide
strong evidence either way.

\section{Conclusions}\label{sec-conc}
We have presented UKIRT $K-$band observations of 222 galaxies selected
from the SDSS first data release based on their strong H$\delta$ absorption, $\ewhd>4$\AA.
The purpose of this paper has been to use $K-$ band luminosities as a
tracer of total stellar mass and morphology to study the connection between
H$\delta$--strong galaxies with (e[a]) and without (k+a) strong nebular emission
(i.e. ongoing star formation).  Our conclusions are the following:

\begin{itemize}
\item The k+a and e(a) galaxies have very different distributions of
  morphology (although there is overlap).  The k+a galaxies are mostly spheroidal; where there is a
  disk, it tends to be smooth with little sign of spiral structure.  On
  the other hand, most e(a) galaxies resemble normal spiral galaxies
  with a prominent disk and spiral structure.  Thus morphology is more
  closely correlated with ongoing star formation (emission lines) than
  with recent star formation (Balmer absorption).

\item The $K-$band fractional bulge luminosities of k+a galaxies are too large to result
  from the truncation of star formation in a typical e(a) or normal spiral galaxy.

\item The k+a galaxies form a surprisingly tight sequence in $(u-g)$
  and $(r-k)$ colour, and H$\delta$ line strength.  A model of an old
  (13.7 Gyr),  dust--free galaxy with a Salpeter initial mass function and a recent burst
  of star formation accounting for 15 per cent of its mass, traces the
  mean properties of this sequence well.  However the burst strength is
  not well constrained; any value $>5$ per cent is consistent within the
  scatter of the data.  The lack of scatter in the dust-sensitive $(r-k)$ colours for
  these galaxies suggests that their unusual spectral appearance is not
  due to the effects of dust.

\item The e(a) galaxies form a sequence in $(u-g)$--$(r-k)$
  colour--colour space that is disjoint from the k+a sequence.
Furthermore, the $(r-k)$ colours of most are too red for their H$\delta$
absorption to be accounted for by the same model that produces k+a galaxies.
Instead, a consistent model for e(a) galaxies is that of a
  dust--reddened (by $\tau_V\sim 2$) but otherwise normal spiral
  galaxy.  However, we have not searched for a unique model and other
  explanations may be admissible, including variations in the initial mass function.

\item With the above interpretation, we can determine the $M/L_K$ ratio
  for each type of galaxy.  While $M/L_K\sim 0.55$ for the e(a)
  population, $M/L_K=0.8\pm0.1$ for the k+a
  galaxies.  This latter value depends on the assumption that the
  starburst accounts for 15 per cent of the stellar mass; a 5 per cent
  burst has only a small effect and the resulting $M/L_K\sim 1$ is not too
  different from normal, evolved stellar populations.

\item Galaxies with strong H$\delta$ and a small amount of emission
  ($\ewoii<10$\AA\ and $\ewha<10$\AA) have colours and morphologies
  similar to those of k+a galaxies, and distinct from e(a) galaxies
  with stronger emission lines.  

\item Both e(a) and k+a galaxies are found in similar environments, as
  characterised by the number of bright neighbouring galaxies.  The
  distribution of environments is typical of SDSS
  galaxies in general, and neither k+a nor e(a) galaxies are
  restricted to cluster cores or outskirts ($<3 R_{\rm vir}$). 

\end{itemize}

We therefore conclude that most e(a) and k+a galaxies are distinct
populations, and arise in different ways.  While most e(a) galaxies appear to
be spiral galaxies with unusually high dust extinction (or perhaps
an atypical initial mass function), most k+a galaxies
are spheroidal galaxies with a recent, substantial ($>5$ per cent by mass) burst
of star formation.  The progenitors of k+a galaxies may still have had
an e(a) spectrum; these could either be the subset of present-day e(a)
galaxies (the bluest examples with evidence of
interactions), or a short-lived blue population that is not represented in
our e(a) sample.  However, most emission line galaxies selected on
H$\delta$ and $r-$magnitude alone are unlikely to be progenitors of k+a
galaxies.

%\acknowledgements
\section*{Acknowledgements}
We thank the observers who carried out our queue observations at UKIRT
in semester 03A: S. Littlefair, Ian Smail, R. Jameson, David Gilbank,
J. Dunlop, Sandy Leggett and Watson Varricatt.
\bibliography{ms}
\newpage
\onecolumn
\setlongtables
\begin{longtable}{cccccccccc}
\caption{The k+a galaxy sample and measured properties\label{tab-ka}}\\
\hline\hline
RA & Dec & z & r$^{\prime}$ &$K_s$ & B/T & $W_\circ(H\delta)$ &
H$\delta_A$ & $W_\circ$(O[{\sc ii}]) & $W_\circ(H\alpha)$\cr
\multispan2{\hfil (J2000)\hfil} & &\multispan2{\hfil (mag)\hfil} & &\multispan4{\hfil (\AA)\hfil} \cr
\hline
\endfirsthead
\multicolumn{10}{l}{Table~\ref{tab-ka} continued}\\
\hline\hline
RA & Dec & z & r$^{\prime}$ &$K_s$ & B/T & $W_\circ(H\delta)$ &
H$\delta_A$ & $W_\circ$(O[{\sc ii}]) & $W_\circ(H\alpha)$\cr
\multispan2{\hfil (J2000)\hfil} & &\multispan2{\hfil (mag)\hfil} & &\multispan4{\hfil (\AA)\hfil} \cr
\hline
\endhead
\hline\multicolumn{10}{r}{\small\sl continued on next page}
\endfoot
\hline
\noalign {Column descriptions: {\it (1-2) }Right ascension and
  declination;{\it (3) }Redshift; {\it (4)} SDSS $r$
  magnitude; {\it (5) } $K_s$ magnitude measured from our UKIRT data.
  Uncertainties are limited by the zeropoint and are $\sim 0.08$
  magnitude;  {\it (6) }The fraction of K-luminosity in the bulge
  component, and uncertainty, as determined by {\sc Gim2D} model
  fits.  Note that these are meaningless in irregular systems (such as
  close pairs or mergers); {\it (7) } rest-frame equivalent width of H$\delta$, in \AA, from
  \citet{tomo-EA1}, where positive numbers represent absorption; {\it
    (8) } rest-frame equivalent width of H$\delta$
  based on the definition of \citet{WO97}; {\it (9-10) } Rest-frame
  equivalent widths of [O{\sc ii}] and H$\alpha$ from the SDSS
  pipeline, measured in \AA\ with positive numbers indicating emission. }\\
\endlastfoot
%\noalign{\smallskip}
3.120312 &$-0.894937$ & 0.128834 & 16.71 & 13.7 & $0.46$ & 4.7 & 4.3 & 5.6 & 5.4\\ 
3.963556 &$-10.388311$ & 0.198220 & 17.5 & 14.7 & $0.67$ & 7.4 & 6 & 1.9 & 0\\ 
9.282578 &$0.410167$ & 0.080655 & 17.11 & 14.3 & $0.35$ & 5.5 & 5.5 & 7.5 & 2.5\\ 
17.773197 &$14.266232$ & 0.099456 & 16.93 & 14.2 & $0.76$ & 5.1 & 4.8 & 3.6 & -\\ 
20.065346 &$-9.988923$ & 0.136923 & 17.22 & 14.5 & $0.56$ & 4.4 & 4.4 & 2 & -\\ 
20.226444 &$13.765402$ & 0.127631 & 16.54 & 13.5 & $0.76$ & 5.5 & 5.2 & 5.9 & 5.3\\ 
26.196381 &$0.537470$ & 0.178903 & 17.46 & 14.4 & $0.42$ & 6.6 & 4.1 & 3 & -\\ 
27.779242 &$-0.943535$ & 0.198065 & 17.38 & 14.5 & $0.7$ & 4.7 & 5.1 & 0.5 & -\\ 
29.191742 &$-9.950396$ & 0.126340 & 17.75 & 14.7 & $0.98$ & 6.1 & 5.9 & 6.8 & 9.2\\ 
30.125355 &$-9.890063$ & 0.095773 & 17.3 & 14.6 & $0.61$ & 4.3 & 4.1 & 2.7 & 2.3\\ 
31.007927 &$14.241354$ & 0.081300 & 17.5 & 14.6 & $0.4$ & 5.3 & 5.2 & 3 & 2.1\\ 
36.112850 &$-9.627632$ & 0.088286 & 17.01 & 14.3 & $0.48$ & 5.2 & 4.8 & 3.9 & -\\ 
36.930046 &$-0.256401$ & 0.219166 & 16.74 & 13.9 & $0.74$ & 5.3 & 5.4 & 2 & 1.2\\ 
37.078602 &$-1.034778$ & 0.091273 & 17.42 & 14.6 & $0.8$ & 5.1 & 4.7 & 7.5 & -\\ 
37.489056 &$-0.903572$ & 0.085923 & 17.18 & 14.2 & $0.43$ & 5.6 & 5.1 & 3.1 & 3.7\\ 
38.693249 &$0.509964$ & 0.140466 & 17.28 & 14.6 & $0.8$ & 6.1 & 5.8 & 2.2 & -\\ 
39.761936 &$-0.530933$ & 0.136121 & 17.45 & 13.8 & $0$ & 4.7 & 1.9 & 7 & -\\ 
45.830013 &$-8.605536$ & 0.075652 & 16.77 & 14.1 & $0.52$ & 4.9 & 5 & 1.1 & -\\ 
46.116833 &$-8.818506$ & 0.119705 & 17.25 & 14.6 & $0.46$ & 6.3 & 6.1 & 1.9 & -\\ 
47.029408 &$0.456219$ & 0.074281 & 15.25 & 12.3 & $0.81$ & 5.9 & 5.9 & 2.8 & 8.5\\ 
51.094597 &$-5.940036$ & 0.204436 & 17.59 & 14.5 & $1$ & 6.4 & 5.5 & 3.8 & 3.3\\ 
60.141998 &$-6.407394$ & 0.179402 & 16.89 & 13.9 & $0.7$ & 5.1 & 3.2 & 1 & -\\ 
113.390167 &$37.834980$ & 0.096724 & 17.45 & 14.6 & $0.36$ & 5.6 & 5.9 & 3.8 & 1.1\\ 
113.905518 &$39.562695$ & 0.108367 & 16.69 & 13.9 & N/A & 5.4 & 5.3 & 1.3 & -\\ 
118.233299 &$38.725178$ & 0.126291 & 17.53 & 14.7 & $0.7$ & 4.7 & 4.1 & 2.9 & -1.1\\ 
124.110924 &$41.562134$ & 0.101673 & 17.29 & 14.2 & $0.6$ & 5.1 & -1.7 & 4.6 & -\\ 
126.852570 &$51.797470$ & 0.081440 & 16.44 & 13.7 & $0.74$ & 4.7 & 4.6 & 1.8 & -\\ 
127.648438 &$49.455341$ & 0.132265 & 16.21 & 13.4 & $0.92$ & 4.7 & 4.6 & 0.7 & -\\ 
128.130447 &$49.411301$ & 0.183051 & 17.24 & 14.4 & $0.43$ & 5.4 & 5.1 & 1.5 & -\\ 
128.838440 &$42.660007$ & 0.091694 & 16.15 & 13.6 & $0.48$ & 6.2 & 6.2 & 2.9 & -\\ 
131.831635 &$3.341675$ & 0.072952 & 16.12 & 13.4 & $0.65$ & 4.6 & 4.6 & 4.5 & -\\ 
135.309357 &$51.391785$ & 0.129297 & 17.32 & 14.3 & $0.56$ & 5.2 & 2.4 & 1.2 & -\\ 
135.702240 &$54.159458$ & 0.101710 & 16.37 & 13.6 & $1$ & 3.8 & 3.5 & 2.9 & -\\ 
135.886566 &$1.210117$ & 0.058030 & 16.13 & 13.2 & $0.37$ & 4.5 & 4.1 & 3.8 & -\\ 
135.887466 &$1.208835$ & 0.057939 & 17.33 & 13.2 & $0.37$ & 4.5 & 2.2 & 1.8 & -\\ 
135.991104 &$55.244137$ & 0.089378 & 17.57 & 14.8 & $0.49$ & 4.6 & 4.1 & 3.1 & -\\ 
136.103760 &$53.937237$ & 0.140552 & 17.61 & 14.7 & $0$ & 6.1 & 5.7 & 3 & -\\ 
136.583054 &$52.363892$ & 0.098701 & 17.52 & 15 & $0.69$ & 6.3 & 6.3 & 1.5 & -\\ 
136.752792 &$0.060915$ & 0.164117 & 17.26 & 14.2 & $0.85$ & 4.9 & 4.8 & 4.6 & 3.6\\ 
138.115753 &$53.706375$ & 0.222322 & 17.3 & 14.1 & $0.93$ & 6.4 & 6.4 & 1.5 & -\\ 
139.567947 &$56.832035$ & 0.115982 & 16.61 & 13.7 & $0.53$ & 4.3 & 2.9 & 1.5 & -\\ 
141.083755 &$3.248529$ & 0.129099 & 16.61 & 13.6 & $0.38$ & 4.5 & 4.3 & 13.8 & -1.4\\ 
144.678787 &$0.030188$ & 0.090723 & 16.07 & 13.1 & $0.47$ & 5 & 5 & 3.9 & -\\ 
145.353928 &$57.963165$ & 0.081976 & 16.44 & 13.7 & $0.94$ & 4.2 & 4.3 & 1.8 & -\\ 
147.077835 &$2.501155$ & 0.060456 & 16.27 & 13.4 & $0.29$ & 6.4 & 6 & 2.4 & -\\ 
148.432892 &$-0.090180$ & 0.083397 & 16.54 & 13.7 & $0.52$ & 5.2 & 5 & 1.8 & 6.1\\ 
149.374603 &$2.828351$ & 0.216094 & 17.33 & 14.4 & $0.78$ & 5.5 & 5.4 & 0.8 & -\\ 
153.439102 &$1.270462$ & 0.105583 & 16.18 & 13 & $0.46$ & 6.8 & 6.7 & 5.2 & 8.2\\ 
153.656876 &$1.157441$ & 0.144184 & 16.8 & 13.9 & $0.84$ & 4.3 & 4.4 & 0.3 & -\\ 
153.832031 &$1.061565$ & 0.215896 & 17.1 & 14.3 & $0.82$ & 6 & 5.9 & 0.6 & -\\ 
154.121719 &$-0.026977$ & 0.104553 & 17.3 & 14.4 & $0$ & 5.5 & 5.3 & 3.6 & 2.7\\ 
160.627304 &$0.578335$ & 0.100022 & 16.63 & 13.9 & $0.7$ & 6.7 & 6.7 & 4.2 & -\\ 
168.387756 &$0.833713$ & 0.152380 & 17.56 & 14.8 & $0.78$ & 5 & 4.6 & 2.3 & -\\ 
168.903183 &$-0.029248$ & 0.143258 & 17.51 & 14.6 & $0.69$ & 4.6 & 2.8 & 3.4 & -\\ 
177.656357 &$1.500157$ & 0.077820 & 16.4 & 13.6 & $0.74$ & 6.6 & 6.8 & 2.9 & -\\ 
181.079483 &$-0.315507$ & 0.093782 & 15.18 & 12.6 & $0.38$ & 7.4 & 7.1 & 1.9 & -\\ 
183.690933 &$-0.919306$ & 0.104813 & 17.18 & 14.1 & $0.16$ & 4.8 & 4.3 & 5.7 & 9.4\\ 
185.485245 &$0.163395$ & 0.106243 & 16.3 & 13.8 & $0.76$ & 7.4 & 7.3 & 2.6 & -\\ 
186.827286 &$-0.407924$ & 0.114163 & 16.98 & 14.3 & $0.46$ & 5.6 & 5.2 & 2.2 & -\\ 
190.720642 &$2.616917$ & 0.084382 & 17.52 & 14.9 & $0.4$ & 6 & 6 & 2 & 1.4\\ 
192.314362 &$-0.630072$ & 0.168910 & 17.49 & 14.2 & $0.17$ & 5.5 & 3.6 & 2.5 & 8.8\\ 
195.686630 &$3.319118$ & 0.068262 & 16.48 & 13.2 & $0.51$ & 4.9 & 4.8 & 7.4 & 8.1\\ 
196.580307 &$-0.982160$ & 0.246145 & 17.29 & 14.1 & N/A & 4.8 & 4.8 & 1.5 & -\\ 
198.945999 &$-0.060723$ & 0.197286 & 17.41 & 14.2 & $0.56$ & 6.5 & 6.1 & 5.8 & 7.2\\ 
199.770584 &$0.519567$ & 0.081329 & 17.56 & 14.8 & $0.9$ & 5.6 & 5.5 & 5 & 1.6\\ 
205.427032 &$2.342579$ & 0.075595 & 16.36 & 13.7 & $0.27$ & 4.9 & 4.6 & 2.2 & -\\ 
207.009094 &$2.068259$ & 0.067782 & 15.91 & 13.3 & $0.67$ & 6 & 6.1 & 2.1 & -\\ 
207.554733 &$2.790413$ & 0.106089 & 16.39 & 13.4 & $0.85$ & 4.4 & 4.3 & 4.5 & 10\\ 
207.628204 &$1.467977$ & 0.072761 & 15.9 & 13.4 & $0.64$ & 6.6 & 6.6 & 3.2 & 1\\ 
209.131607 &$-0.268947$ & 0.163201 & 17.26 & 14.4 & $0.55$ & 4.2 & 4.4 & 3.9 & 1.1\\ 
213.135956 &$0.098394$ & 0.126709 & 17.39 & 14.2 & $0.06$ & 5.3 & -1.3 & - & 4.2\\ 
214.725433 &$5.096523$ & 0.079549 & 16.35 & 13 & $0.54$ & 6.2 & 6.1 & 6.3 & 2.5\\ 
217.917770 &$3.031690$ & 0.152753 & 17.66 & 14.7 & $0.52$ & 4.4 & 4.2 & 1.1 & -\\ 
218.522186 &$4.967597$ & 0.087605 & 17.56 & 15 & $0.63$ & 5.9 & 6.1 & 3.2 & -\\ 
218.558441 &$3.027856$ & 0.301874 & 17.71 & 14.7 & $0.44$ & 5.8 & 5.3 & 1.4 & 1.6\\ 
218.704910 &$-0.155541$ & 0.129915 & 17.66 & 14.6 & $0.56$ & 6.4 & -0.3 & 2.3 & -\\ 
218.929077 &$0.948110$ & 0.120438 & 17.27 & 14.5 & $0.9$ & 6.1 & 5.4 & 3.8 & 1.2\\ 
219.054184 &$3.475608$ & 0.119374 & 16.95 & 14 & $0.43$ & 4.2 & 4.1 & 6.2 & 10.1\\ 
219.671616 &$0.173685$ & 0.105131 & 17.14 & 14.4 & $0.78$ & 4.9 & 4.6 & 4.1 & -\\ 
221.547577 &$-0.238415$ & 0.080029 & 17.33 & 17.8 & $0.03$ & 5 & 4.3 & 3.7 & 9.5\\ 
222.647125 &$3.744416$ & 0.116050 & 17.66 & 13.4 & $0.52$ & 5.9 & 5.5 & 5.2 & 4.1\\ 
222.891068 &$2.449617$ & 0.096892 & 17.16 & 14.5 & $0.59$ & 4.7 & 4.8 & 1.8 & -\\ 
223.189453 &$2.889261$ & 0.264362 & 17.74 & 14.8 & $0.63$ & 6.8 & 6.6 & 1.5 & -\\ 
237.330612 &$57.389172$ & 0.225225 & 17.35 & 14.4 & $0.72$ & 5.3 & 4.9 & 0.4 & -0.8\\ 
238.179855 &$54.813190$ & 0.204227 & 17.65 & 14.5 & $0.44$ & 5.1 & 1.6 & 0.4 & -0.7\\ 
238.676956 &$54.844582$ & 0.094013 & 17.62 & 14.7 & N/A & 5.4 & 2.2 & 8.4 & 1.1\\ 
239.025391 &$51.711884$ & 0.123737 & 17.07 & 14.3 & $0.55$ & 4.9 & 4.1 & 2.1 & -1.3\\ 
240.205154 &$54.298256$ & 0.064590 & 16.47 & 13.6 & $0.21$ & 7.4 & 6.7 & - & 6.9\\ 
240.544449 &$2.751429$ & 0.244081 & 17.36 & 14.5 & $0.45$ & 6.2 & 6.2 & 1.9 & -1.5\\ 
242.703766 &$50.720726$ & 0.104457 & 16.77 & 14 & $0.8$ & 5.6 & 5.5 & 2.9 & 1.1\\ 
246.983154 &$48.014202$ & 0.125929 & 17.49 & 14.7 & $0.73$ & 5.2 & 5.1 & 0.8 & -\\ 
247.999588 &$43.298462$ & 0.225063 & 17.67 & 14.5 & $0.75$ & 4.6 & 3.5 & 1.2 & -\\ 
249.646042 &$44.063736$ & 0.112326 & 17.45 & 14.5 & $0.56$ & 6.7 & 6.8 & 2.8 & 7.6\\ 
250.729843 &$41.893227$ & 0.072600 & 16.15 & 13.4 & $0.64$ & 6.5 & 6.7 & 1.2 & -\\ 
251.366440 &$45.754738$ & 0.154507 & 17.51 & 14.7 & $0.48$ & 4.4 & 4.3 & 0.9 & -1.2\\ 
252.795532 &$37.726040$ & 0.136950 & 17.4 & 14.8 & $0.08$ & 7.2 & 7.2 & 9.4 & 2.5\\ 
256.673676 &$38.576031$ & 0.125296 & 17.45 & 14.6 & $0.71$ & 5.5 & 5.3 & 6.5 & 1.9\\ 
258.942749 &$58.382076$ & 0.127103 & 17.42 & 14.7 & $0.48$ & 6.5 & 5.3 & 2.2 & 1.3\\ 
262.832245 &$55.128433$ & 0.102557 & 16.95 & 14.3 & $0.64$ & 5.7 & 5 & 5.2 & -\\ 
263.207581 &$54.367542$ & 0.161453 & 17.37 & 14.8 & $0.64$ & 7 & 6.7 & 4.9 & 7.3\\ 
265.387665 &$54.057060$ & 0.113601 & 17.41 & 14.7 & $0.79$ & 5.7 & 6.2 & 3.7 & -\\ 
309.355164 &$-6.660254$ & 0.114976 & 17.15 & 14.1 & $0.72$ & 4.5 & 4.1 & 2.3 & -\\ 
310.385773 &$-5.226147$ & 0.062059 & 16.73 & 14.1 & $0.62$ & 5.3 & 4.8 & 2.9 & 1.9\\ 
315.745331 &$10.550177$ & 0.092800 & 15.22 & 12.5 & $0.58$ & 4.7 & 4.4 & 1.3 & -\\ 
318.271515 &$-7.736905$ & 0.088147 & 17.62 & 14.9 & $0.59$ & 6.1 & 5.6 & 5.2 & -\\ 
320.108154 &$-6.913008$ & 0.224315 & 17.42 & 14.6 & $0.63$ & 5.2 & 5 & 9 & 5.1\\ 
320.151672 &$-6.622945$ & 0.119113 & 17.54 & 14.8 & $0.76$ & 5 & 4.1 & 2.7 & -\\ 
326.767517 &$10.954961$ & 0.083870 & 16.53 & 13.9 & $0.35$ & 4.4 & 4.2 & 4.3 & -\\ 
327.472046 &$0.969026$ & 0.172979 & 17.47 & 14.3 & $0.55$ & 5.5 & 2.6 & 13.9 & -0.4\\ 
327.512482 &$-7.130242$ & 0.083176 & 17.53 & 14.8 & $0.22$ & 6.1 & 4.1 & 4 & 9.4\\ 
327.862793 &$12.987449$ & 0.206182 & 17.36 & 14.6 & $0.46$ & 6.6 & 6.5 & 9.9 & 2.3\\ 
328.255249 &$-7.838636$ & 0.073321 & 17.26 & 14.2 & $0.38$ & 5.3 & 1.6 & 3.7 & -\\ 
328.746399 &$-6.853999$ & 0.216543 & 17.09 & 14.1 & $0.7$ & 6.3 & 5.9 & 2.4 & -\\ 
329.933380 &$12.793406$ & 0.122527 & 16.47 & 13.6 & $0.56$ & 5.8 & 5.7 & 5.7 & 5.8\\ 
331.439728 &$-9.487941$ & 0.135392 & 17.71 & 15 & $0.43$ & 6 & 5.9 & 3.2 & -\\ 
340.961884 &$-10.229675$ & 0.085119 & 17.63 & 14.9 & $0.29$ & 5.1 & 4.7 & 8.2 & 1.2\\ 
341.995117 &$-9.762263$ & 0.177042 & 17.69 & 14.6 & $0.61$ & 5 & 0.8 & -6.2 & 0\\ 
344.118774 &$14.673409$ & 0.163308 & 17.09 & 14.2 & $0.64$ & 5.1 & 4.6 & 6.3 & -\\ 
344.236511 &$13.067387$ & 0.209131 & 17.18 & 14.4 & $0.72$ & 6.7 & 6.6 & 2.4 & -\\ 
345.029846 &$15.317498$ & 0.082219 & 16.89 & 14 & $0.56$ & 6.1 & 5.8 & 3.8 & 1.7\\ 
345.367645 &$-8.565503$ & 0.181151 & 17.36 & 14.4 & $0.53$ & 5.1 & 4.9 & 3 & 0.9\\ 
346.157288 &$14.149549$ & 0.122222 & 16.53 & 13.7 & $0.48$ & 4.6 & 4.4 & 3.8 & 2.6\\ 
346.930847 &$15.432889$ & 0.069696 & 16.17 & 13.8 & $0.45$ & 6.3 & 6.3 & 1.7 & -\\ 
348.280731 &$15.481363$ & 0.109839 & 17.03 & 13.8 & $0.44$ & 5 & 5.1 & 7.2 & 7.1\\ 
349.948700 &$0.721130$ & 0.118550 & 17.28 & 14.5 & $0.72$ & 5 & 4.4 & 3.8 & -\\ 
353.721649 &$14.846871$ & 0.064671 & 15.98 & 13.4 & $0.56$ & 4.9 & 4.5 & 1.6 & -\\ 
355.495483 &$14.908378$ & 0.085547 & 16.73 & 13.9 & $0.6$ & 5.5 & 5.9 & 6.6 & 2.6\\ 
358.630005 &$14.332058$ & 0.078535 & 15.69 & 12.9 & $1$ & 3.9 & 4 & 4.4 & 1.4\\ 

\end{longtable}
%\end{table*}

\begin{longtable}{cccccccccc}
\caption{The e(a) galaxy sample and measured properties\label{tab-ea}}\\
\hline\hline
RA & Dec & z & r$^{\prime}$ &$K_s$ & B/T & $W_\circ(H\delta)$ &
H$\delta_A$ & $W_\circ$(O[{\sc ii}]) & $W_\circ(H\alpha)$\cr
\multispan2{\hfil (J2000)\hfil} & &\multispan2{\hfil (mag)\hfil} & &\multispan4{\hfil (\AA)\hfil} \cr
\hline
\endfirsthead
\multicolumn{10}{l}{Table~\ref{tab-ea} continued}\\
\hline\hline
RA & Dec & z & r$^{\prime}$ &$K_s$ & B/T & $W_\circ(H\delta)$ &
H$\delta_A$ & $W_\circ$(O[{\sc ii}]) & $W_\circ(H\alpha)$\cr
\multispan2{\hfil (J2000)\hfil} & &\multispan2{\hfil (mag)\hfil} & &\multispan4{\hfil (\AA)\hfil} \cr
\hline
\endhead
\hline\multicolumn{10}{r}{\small\sl continued on next page}
\endfoot
\hline
\endlastfoot
%\noalign{\smallskip}
0.304384 &$-11.103624$ & 0.119084 & 16.44 & 13.4 & $0.31$ & 5 & 4.8 & 19.9 & 47.3\\ 
1.336491 &$-10.469523$ & 0.076842 & 16.7 & 14.1 & $0.01$ & 7.2 & 4.8 & - & 40.6\\ 
2.689777 &$-0.127752$ & 0.086695 & 16.53 & 13.7 & $0.09$ & 5.3 & 5.1 & 13.5 & 32.8\\ 
4.726562 &$-11.208706$ & 0.060786 & 16 & 13.8 & $0.49$ & 7.1 & 4.7 & 29.9 & 51.7\\ 
5.937845 &$-1.008309$ & 0.066280 & 17.14 & 13.9 & $0.33$ & 5.7 & 5.3 & 10.2 & 14.4\\ 
7.592993 &$15.013621$ & 0.098497 & 16.15 & 13.4 & $0.08$ & 6 & 4.6 & 10.3 & 31.6\\ 
10.653961 &$-1.029383$ & 0.252882 & 18.52 & 15.4 & $0.57$ & 5.4 & 2.9 & 16.4 & 37.8\\ 
10.851816 &$0.016397$ & 0.081329 & 16.09 & 13.5 & $0.12$ & 7 & 4.1 & 19.5 & 67.5\\ 
14.807940 &$0.794905$ & 0.172719 & 18.02 & 15.1 & $0.23$ & 5.9 & 4 & 29.7 & 47.9\\ 
15.323996 &$0.524027$ & 0.145131 & 17.49 & 14.6 & $0.2$ & 6.2 & 4.6 & 11.6 & 35.6\\ 
16.047445 &$-10.786201$ & 0.082317 & 17.68 & 15 & $0.44$ & 6.1 & 5.9 & 25.5 & 15.3\\ 
16.591736 &$-0.535420$ & 0.171226 & 17.3 & 14 & $0.61$ & 4.6 & 4.4 & 30.7 & 34.9\\ 
17.969322 &$-9.766603$ & 0.131466 & 17.35 & 13.9 & $0.61$ & 5.1 & 5.1 & 35.5 & 31.7\\ 
21.064075 &$14.038528$ & 0.103083 & 17.4 & 13.9 & $0.15$ & 5.9 & 3.5 & 4.3 & 20.8\\ 
21.218010 &$-9.570095$ & 0.074324 & 16.78 & 13.6 & $0.79$ & 5.2 & 5.3 & 18.8 & 8.1\\ 
24.083879 &$14.578506$ & 0.107873 & 17.41 & 14.6 & $0.16$ & 5.1 & 4.6 & 9.5 & 37\\ 
24.298273 &$13.111508$ & 0.054026 & 16.28 & 13.8 & $0.12$ & 5.4 & 5 & 22.1 & 28.8\\ 
24.492804 &$-8.575119$ & 0.189068 & 17.39 & 13.5 & $0.88$ & 4.7 & 4.6 & 15.3 & 32.2\\ 
27.441174 &$0.808579$ & 0.162652 & 17.31 & 14.3 & $0.29$ & 5 & 4.9 & 8.3 & 25.6\\ 
29.266397 &$13.594388$ & 0.100934 & 16.43 & 13.5 & $0.64$ & 4.5 & 4.5 & - & 34.6\\ 
29.476841 &$14.095569$ & 0.178564 & 17.49 & 14.5 & $0$ & 5.1 & 4.5 & 8.8 & 37.5\\ 
35.197891 &$-9.624654$ & 0.108725 & 16.85 & 13.6 & $0.57$ & 4.7 & 4.5 & 17.3 & 17.6\\ 
40.521065 &$-8.866433$ & 0.110401 & 16.04 & 13.1 & $0.24$ & 4.5 & 4.3 & 8.9 & 34.9\\ 
41.083641 &$-7.754149$ & 0.076250 & 15.55 & 12.9 & $0.71$ & 4.6 & 4.7 & 22.2 & 39.2\\ 
46.296959 &$-6.431061$ & 0.086905 & 17.53 & 15.3 & $0$ & 5.9 & 6 & 10.9 & 15.9\\ 
52.038315 &$-7.305844$ & 0.086911 & 17.64 & 15.2 & $0.05$ & 5.2 & 5.2 & - & 32.7\\ 
62.729263 &$-6.176466$ & 0.129256 & 17.35 & 14.1 & $0.34$ & 5.5 & 4.6 & 19.3 & 34\\ 
116.479462 &$32.579166$ & 0.100922 & 17.21 & 14.3 & $0.06$ & 5 & 5 & 6.5 & 16.3\\ 
117.033798 &$42.903248$ & 0.113352 & 17.66 & 14.6 & $0.28$ & 5.4 & 4 & 8.1 & 21\\ 
119.685112 &$40.516521$ & 0.074947 & 16.72 & 13.8 & $0.44$ & 5.5 & 4.7 & 4 & 16.5\\ 
123.469513 &$40.901772$ & 0.100391 & 17.33 & 14.1 & $0.07$ & 5 & 2.6 & 5.4 & 12.6\\ 
130.364212 &$3.535027$ & 0.143611 & 17.44 & 14.4 & $0.67$ & 4.7 & 4.5 & 2.9 & 13.6\\ 
133.691208 &$48.470387$ & 0.119963 & 17.41 & 14.2 & $0.03$ & 4.7 & 2.3 & 4.6 & 19.1\\ 
148.434738 &$-0.090464$ & 0.083706 & 15.64 & 13.7 & $0.52$ & 5.2 & 3 & 13.3 & 33.3\\ 
148.655762 &$2.207545$ & 0.162581 & 17.25 & 14.1 & $0.76$ & 5.4 & 4.1 & 6 & 19\\ 
163.710587 &$-0.982576$ & 0.074838 & 17.55 & 14.8 & $0.23$ & 5.2 & 4.8 & 7.2 & 16.3\\ 
166.723526 &$1.249782$ & 0.105472 & 16.09 & 12.9 & $0.36$ & 4.6 & 4.1 & 4.9 & 14.8\\ 
173.176758 &$0.756327$ & 0.067486 & 16.83 & 12.8 & $0.71$ & 5.5 & 6.1 & 19.7 & 26\\ 
175.077286 &$1.076113$ & 0.076980 & 16.6 & 13 & $0.71$ & 10.5 & 8.3 & 9.7 & 19.7\\ 
189.067123 &$-0.386185$ & 0.090503 & 17.29 & 14.6 & $0.86$ & 5.6 & 5.9 & 11.6 & 1.8\\ 
197.523224 &$-2.472836$ & 0.132543 & 17.52 & 14.6 & $0.13$ & 4.6 & 4.8 & 10.4 & 12.1\\ 
198.926941 &$-1.724550$ & 0.175708 & 17.48 & 14.3 & $0.04$ & 6.8 & 2.4 & 5.1 & 18.4\\ 
201.826462 &$0.865157$ & 0.111664 & 17.18 & 14.3 & $0.31$ & 5.3 & 4.3 & 5.4 & 18.7\\ 
210.408279 &$0.067353$ & 0.085131 & 15.66 & 13.1 & $0.18$ & 4.3 & 4 & 5.8 & 15\\ 
214.608963 &$1.356086$ & 0.083081 & 16.46 & 13.6 & $0.28$ & 5.2 & 4 & 3.3 & 16.8\\ 
222.645447 &$3.745237$ & 0.115040 & 16.8 & 13.4 & $0.52$ & 5.9 & 0.1 & 42.8 & 78.8\\ 
239.997879 &$54.588268$ & 0.057877 & 16.63 & 13.7 & $0.01$ & 5.7 & 3.1 & 5.4 & 20.4\\ 
240.273132 &$49.979671$ & 0.071001 & 15.43 & 12.6 & $0.25$ & 6 & 5.6 & 30.7 & 34.4\\ 
241.539871 &$51.154034$ & 0.077398 & 17.19 & 14 & $0$ & 7.4 & 2.6 & 8.5 & 18.6\\ 
246.701721 &$45.584122$ & 0.086421 & 17.24 & 14.3 & $0.13$ & 7.6 & 5 & 13.6 & 28.5\\ 
247.959869 &$44.898746$ & 0.222740 & 17.28 & 14.3 & $0.3$ & 4.2 & 4 & 28.6 & 37.3\\ 
248.377213 &$43.259373$ & 0.122890 & 16.73 & 14.4 & $0.21$ & 7.2 & 5 & 26.4 & 58.3\\ 
252.343353 &$38.528667$ & 0.086168 & 16.32 & 13.9 & $0.05$ & 5.9 & 4.7 & 19.2 & 41\\ 
253.869644 &$40.062359$ & 0.061714 & 16.98 & 14.7 & $0$ & 4.9 & 4.8 & 33.2 & 46.9\\ 
254.030594 &$38.869167$ & 0.073432 & 17.17 & 13.9 & $0.01$ & 6.6 & 5 & 13.5 & 26.3\\ 
254.062775 &$38.859486$ & 0.051833 & 15.95 & 12.6 & $0.47$ & 5.1 & 4.8 & 15.4 & 27.6\\ 
258.726166 &$55.708164$ & 0.106540 & 16.74 & 13.7 & $0.68$ & 6.2 & 4.8 & 10.4 & 25.3\\ 
260.090363 &$54.825855$ & 0.100796 & 17.57 & 14.3 & $0.58$ & 6 & 4.1 & 8 & 20.5\\ 
260.632111 &$52.869144$ & 0.159600 & 17.03 & 14.2 & $0.36$ & 6.2 & 4.3 & 14.6 & 38.3\\ 
262.566376 &$54.298878$ & 0.085144 & 16.93 & 14.2 & $0.2$ & 4.6 & 4.3 & 4.7 & 16.8\\ 
264.509949 &$55.871490$ & 0.084523 & 16.73 & 13.8 & $0.12$ & 5.7 & 4.3 & 15.3 & 31.3\\ 
309.668823 &$-4.640462$ & 0.229230 & 17.3 & 14.1 & $0.1$ & 6.6 & 4.8 & 5 & 23.6\\ 
309.947723 &$-6.421526$ & 0.111621 & 17.36 & 14.7 & $0.18$ & 5 & 4.5 & 25.6 & 39.3\\ 
316.396729 &$-6.442801$ & 0.086264 & 17.2 & 13.8 & $0.28$ & 5.8 & 5.2 & 15.7 & 15\\ 
317.146118 &$-8.015185$ & 0.122152 & 16.71 & 13.2 & $0.52$ & 4.8 & 5 & 4.1 & 14.7\\ 
318.084625 &$-7.201864$ & 0.098725 & 16.63 & 13.2 & $0.39$ & 6 & 5.6 & 4508.6 & 2.7\\ 
322.490906 &$-8.057539$ & 0.219952 & 17.25 & 14.4 & $0.69$ & 4.3 & 4.3 & 4.7 & 24.5\\ 
322.521881 &$-6.686210$ & 0.089172 & 14.52 & 15.2 & $0.32$ & 5 & 3.6 & 91.5 & 38.4\\ 
323.297852 &$-8.495975$ & 0.091285 & 17.29 & 14.3 & $0.5$ & 5.2 & 3.3 & 7.8 & 15.5\\ 
323.427399 &$-8.360456$ & 0.117046 & 16.4 & 13.4 & $0.64$ & 6.7 & 5.7 & 10.8 & 28.4\\ 
327.134552 &$0.133313$ & 0.158565 & 17.64 & 14.5 & $0.46$ & 5.4 & 4.6 & 7.1 & 14\\ 
327.349152 &$-8.675175$ & 0.102440 & 17.14 & 13.7 & $0.93$ & 5 & 3.8 & 17 & 30.6\\ 
327.423340 &$12.339659$ & 0.076621 & 16.81 & 13.9 & $0.28$ & 4.8 & 5 & 5.4 & 15.5\\ 
327.666290 &$-1.073946$ & 0.103408 & 17.23 & 14.7 & $0.26$ & 5.3 & 4.9 & 19.1 & 29.9\\ 
327.735779 &$-6.819699$ & 0.058740 & 15.98 & 13.3 & $0.8$ & 4.7 & 5.2 & 24.6 & 13.9\\ 
327.868164 &$-6.775196$ & 0.101240 & 16.67 & 13.9 & $0.95$ & 7.4 & 6.9 & 16 & 19.6\\ 
330.465393 &$12.099227$ & 0.123104 & 16.06 & 13.1 & $0.55$ & 5.8 & 5.5 & 26.1 & 35\\ 
330.580353 &$-0.565185$ & 0.084469 & 17.62 & 12.9 & $0.09$ & 7 & 7.2 & 14.4 & 24.1\\ 
334.127258 &$-0.574393$ & 0.111474 & 16.73 & 14 & $0.87$ & 6.4 & 4.7 & 25.1 & 38.9\\ 
334.428925 &$-0.290766$ & 0.094682 & 16.96 & 14.1 & $0.36$ & 5.6 & 4.6 & 25.8 & 47.4\\ 
334.805664 &$12.976340$ & 0.106792 & 17.31 & 14.4 & $0.4$ & 5.1 & 4.8 & 20 & 44.3\\ 
336.138947 &$0.579151$ & 0.090261 & 18.97 & 16.6 & $0.08$ & 5.7 & 5.5 & 16 & 29.1\\ 
340.294891 &$-9.014417$ & 0.071547 & 15.59 & 12.3 & $0.22$ & 5.4 & 4.6 & 15.4 & 34.9\\ 
340.507233 &$-8.774342$ & 0.166393 & 17.76 & 13.6 & $0.28$ & 5.6 & 5.3 & 15.4 & 11.7\\ 
340.507812 &$-8.773771$ & 0.166596 & 17.64 & 13.6 & $0.28$ & 5.6 & 5.4 & 11.6 & 4.9\\ 
342.494751 &$0.918355$ & 0.089213 & 16.47 & 13.9 & $0.06$ & 5.1 & 5.3 & 15.3 & 35.4\\ 
342.842072 &$-8.956442$ & 0.080072 & 17.22 & 14.4 & $0.54$ & 5.8 & 5.9 & 33.6 & 21.9\\ 
346.197815 &$14.275584$ & 0.082477 & 17.02 & 14.7 & $0.11$ & 7.5 & 7.2 & 22.5 & 28.9\\ 
346.333771 &$-9.530315$ & 0.183999 & 17.38 & 14.2 & $0.55$ & 5.4 & 5.5 & 4.4 & 12.3\\ 
346.764313 &$-10.351763$ & 0.104952 & 17.13 & 14.2 & $0.12$ & 4.9 & 5 & 14 & 16.1\\ 
348.372345 &$0.909661$ & 0.120993 & 17.36 & 14.3 & $0.22$ & 6.9 & 2.1 & 8.1 & 12.1\\ 
356.560333 &$13.656058$ & 0.077711 & 17.1 & 14.2 & $0.12$ & 6.5 & 4.6 & 9.9 & 26.7\\ 
358.733002 &$-10.032640$ & 0.111582 & 17.27 & 14.3 & $0.64$ & 5.6 & 4.6 & 33.5 & 84.8\\ 
358.946411 &$16.096483$ & 0.102185 & 17.66 & 14.3 & $0.18$ & 6.2 & 5.7 & 10.8 & 18.3\\ 

\end{longtable}
\newpage
\appendix
\section{Galaxy Morphologies}\label{sec-morph}
In this appendix we present images for a subset of our sample.  
Figures~\ref{fig-EAtrue_disk}--\ref{fig-EAtrue_bulge} show all of the k+a galaxies, 
sorted by B/T ratio.
Figures~\ref{fig-EAweakem_disk}--\ref{fig-EAweakem_bulge} show the same for a
representative sample of k+a--like galaxies with weak emission lines
($\ewha<10$\AA\ and $\ewoii<10$\AA).  There are another nine of
these galaxies in our sample, not shown here: all but one have $B/T>0.5$.
Finally, in Figures~\ref{fig-EAem_blue}--\ref{fig-EAem_bulge} we show
images for 60 of the 94 e(a) galaxies;  Figure~\ref{fig-EAem_blue} shows the twelve bluest
e(a) galaxies, while Figures~\ref{fig-EAem_disk} and
\ref{fig-EAem_bulge} show a representative subsample of the remaining
e(a) galaxies with disk--dominated morphologies or
bulge--dominated morphologies, respectively.  
In each panel we label the B/T
measurement and the galaxy ID number.  A star is shown in the
upper--left corner of the image if we have identified it as possibly
interacting.  
\suppressfloats

\begin{figure}
%    \center\leavevmode\epsfxsize=15cm%\epsfbox{FigA1.epsi}
\caption{$K-$ band images and model fits for k+a galaxies with
  $B/T\leq0.52$.  In each set of three images, the left image is the
  central 12\arcsec\ of the original image; the middle panel shows the
  {\sc GIM2D} model fit; and the right panel shows the residual between
  the model and the data.  Contours shown on the model fits are logarithmically spaced.  The
  $B/T$ value and galaxy id are given above each set of three images.  A star is shown in the
upper--left corner of the image if we have identified it as possibly
interacting; note that this identification is often based on features
that are either difficult to discern on these small images, or are
outside the image boundaries.  
\label{fig-EAtrue_disk}}
\end{figure}
%\newpage

\begin{figure}
%\leavevmode \epsfxsize=18cm \epsfbox{FigA2.epsi}
\caption{As Figure~\ref{fig-EAtrue_disk}, but for k+a galaxies with
  $0.52< B/T<0.7$.
\label{fig-EAtrue_int}}
\end{figure}

%\newpage
\begin{figure}
%\leavevmode \epsfxsize=18cm \epsfbox{FigA3.epsi}
\caption{As Figure~\ref{fig-EAtrue_disk}, but for k+a galaxies with
  $B/T\geq 0.7$.
\label{fig-EAtrue_bulge}}
\end{figure}

%\newpage
\begin{figure}
%\leavevmode \epsfxsize=18cm \epsfbox{FigA4.epsi}
\caption{As Figure~\ref{fig-EAtrue_disk}, but for a representative
  subsmple of k+a galaxies with a
  small amount of emission ($\ewha<10$\AA\ and $\ewoii<10$\AA) and
  $B/T<0.5$.  There is one other galaxy in our sample satisfying these
  criteria, but not shown.
\label{fig-EAweakem_disk}}
\end{figure}
%\newpage

\begin{figure}
%\leavevmode \epsfxsize=18cm \epsfbox{FigA5.epsi}
\caption{As Figure~\ref{fig-EAweakem_disk}, but for $B/T>0.5$.  There
  are an additional 8 galaxies in this category, not shown here.
\label{fig-EAweakem_bulge}}
\end{figure}

%\newpage
\begin{figure}
%\leavevmode \epsfxsize=18cm \epsfbox{FigA6.epsi}
\caption{As Figure~\ref{fig-EAtrue_disk}, but for the bluest e(a)
  galaxies, with $(u-g)<1.1$ and $(r-k)<2.7$.
\label{fig-EAem_blue}}
\end{figure}

%\newpage
\begin{figure}
%\leavevmode \epsfxsize=18cm \epsfbox{FigA7.epsi}
\caption{As Figure~\ref{fig-EAtrue_disk}, but for a representative
  sample e(a) galaxies with
  $B/T<0.5$, excluding those shown in Figure~\ref{fig-EAem_blue}.  There are another 31 galaxies of this type not shown.
\label{fig-EAem_disk}}
\end{figure}

%\newpage
\begin{figure}
%\leavevmode \epsfxsize=18cm \epsfbox{FigA8.epsi}
\caption{As Figure~\ref{fig-EAem_disk}, but for $B/T>0.5$, and excluding those shown in Figure~\ref{fig-EAem_blue}.  There are
  another 3 galaxies of this type, not shown.
\label{fig-EAem_bulge}}
\end{figure}

\end{document}